\documentclass[traditabstract]{aa}
\usepackage{aalongtable}
\usepackage{natbib,twoopt}
\bibpunct{(}{)}{;}{a}{}{,}   
\usepackage{graphics}
\usepackage{ulem}
\usepackage{times}
\usepackage{psfig}
\begin{document}
\title{The Mg {\sc ii} $\lambda$2797, $\lambda$2803 emission
in low-metallicity star-forming galaxies from the SDSS\thanks{Table \ref{tab2} 
is only available in the electronic edition.}}
\author{N. G. \ Guseva \inst{1,2}
\and Y. I. \ Izotov \inst{1,2}
\and K. J. \ Fricke \inst{1,3}
\and C. \ Henkel \inst{1,4}
}
\offprints{N. G. Guseva, guseva@mao.kiev.ua}
\institute{          Max-Planck-Institut f\"ur Radioastronomie, Auf dem H\"ugel 
                     69, 53121 Bonn, Germany
\and
                     Main Astronomical Observatory,
                     Ukrainian National Academy of Sciences,
                     Zabolotnoho 27, Kyiv 03680,  Ukraine
\and
                     Institut f\"ur Astrophysik, G\"ottingen Universit\"at, 
                     Friedrich-Hund-Platz 1, 37077 G\"ottingen, Germany 
\and
                     Astronomy Department, King Abdulaziz University, 
                     P.O. Box 80203, Jeddah,  Saudi Arabia
}
\date{Received \hskip 2cm; Accepted}

\abstract{We present 65 Sloan Digital Sky Survey (SDSS) spectra of 62 
star-forming galaxies 
with oxygen abundances 12 + logO/H $\sim$ 7.5 -- 8.4. Redshifts 
of selected galaxies are in the range
$z$ $\sim$ 0.36 -- 0.70. This allows us to detect the redshifted 
 Mg~{\sc ii} $\lambda$2797,  $\lambda$2803 emission lines.
Our aim is to use these lines for the magnesium abundance determination. 
The Mg~{\sc ii} emission was detected in $\sim$ 2/3 of the
galaxies. We find that the Mg {\sc ii} $\lambda$2797 emission-line 
intensity follows a trend with the excitation parameter 
$x$ = O$^{2+}$/O that is similar
to that predicted by CLOUDY photoionised H~{\sc ii} region models,
suggesting a nebular origin of Mg {\sc ii} emission. The Mg/O abundance ratio
is lower by a factor $\sim$ 2 than the solar ratio. This is probably the 
combined effect of interstellar Mg~{\sc ii} absorption and
depletion of Mg onto dust. However, the effect of dust depletion in selected
galaxies, if present, is small, by a factor of $\sim$2 lower than that of
 iron.
}
\titlerunning{The Mg {\sc ii} $\lambda$2797, $\lambda$2803 emission
in star-forming galaxies from the SDSS}
\authorrunning{N. G. Guseva et al.}
\maketitle

\section {Introduction}\label{S1}

Magnesium is an $\alpha$-process element produced during nuclear
burning in massive stars, which is similar to oxygen and some other elements, 
such as noble neon and argon. Its abundance is studied in detail from
absorption lines in stars with a wide range of metallicities and shows trends 
similar to oxygen. 

   The Mg {\sc ii} $\lambda$$\lambda$ 2797,2803\AA\ doublet (h and k) 
is one of the
most studied tracers of the gas-phase medium in planetary nebulae (PNe),
 in the Local InterStellar Medium  (LISM)
of the Galaxy, and in the gaseous environment of distant galaxies. 
 Since magnesium, which is similar
to silicon and iron among other elements, is a refractory element, 
its abundance can 
provide information
about the level of interstellar magnesium depletion onto dust.

 The magnesium abundance in the LISM 
can be determined by analysing absorption lines towards Galactic stars.
  With high-resolution spectra of many Galactic early-type stars 
observed from \textit{Copernicus} launched in 1972 
it was found that
profiles of Mg {\sc ii} h and k lines are identical with solar profiles
except for the presence of narrow absorption components formed 
in the interstellar medium
along the line of sight \citep{Oegerle,Murray1983}. 
 Follow-up observations of early type and cool stars with the
\textit{IUE (International Ultraviolet Explorer)} and the
\textit{HST (Hubble Space Telescope)} 
provided quantitative characteristics of the interstellar medium   towards 
Galactic stars and planetary nebulae: the distribution of the magnesium 
abundance in the LISM \citep{Molaro1986,M88},
specifically; the structure of the LISM within 100 pc \citep{Redfield2004}; 
and the abundances and physical conditions in the diffuse interstellar clouds
\citep{Welty1999}. These data indicate a moderate
level of magnesium depletion.

There are only a few determinations of the magnesium abundance in 
H {\sc ii} regions, such as the Orion nebula in the Galaxy and the 
Tarantula nebula in the Large Magellanic Cloud 
and in some bright planetary nebulae \citep{Rodriguez2005,PP10,Dinerstein2012}.
   \citet{Dinerstein2012} has investigated the gas-phase 
magnesium abundances in 25 planetary nebulae using the 
Mg~{\sc ii} $\lambda$4481\AA\ recombination line.
  They find that Mg/H is 
close to solar, implying that Mg is at most 
minimally depleted, whereas the measurements in the 30 Doradus and 
the Orion nebulae,
with the same recombination line,  
indicate significantly higher levels of Mg depletion up to 72\%  and 90\%, 
respectively \citep{PP10}.

 The Mg {\sc ii} $\lambda$$\lambda$2707,2803\AA\ doublet is the 
strongest absorption 
feature that is detectable in the optical range at intermediate redshifts 
(0.3 $\leq$ $z$ $\leq$ 2). Therefore, magnesium
absorption lines are often detected
in spectra of background quasars and galaxies.
  Spectra of background sources allow us to probe regions of 
low-ionisation and cool gas 
present in the outer regions of spiral discs and dwarf irregular galaxies
\citep{DZavadsky2004}.
 Many Mg {\sc ii} surveys have been carried out for these systems 
\citep{Sargent1988,Steidel1992,DZavadsky2003,York2006,Prochter2006,Prochaska2007,Lilly2007,Quider2011}.  
Owing to large spectroscopic surveys including 
the Sloan Digital Sky Survey (SDSS)
\citep{Y00} hyndreds of thousands of quasars have been 
used to identify large samples of absorption systems along the line 
of sight using the Mg {\sc ii} absorption lines 
\citep{York2006,Bouche2006,Prochter2006,Lundgren2009,Quider2011}.
   Recently, based on a fully-automatic method, \citet{Zhu2012} compiled 
a very large sample 
of $\sim$ 40,000 Mg {\sc ii} absorbers 
from the SDSS Data Release 7 (DR7). 

   Using the Mg {\sc ii} absorber samples many detailed studies on the 
distribution of column densities, the redshift evolution of densities,
and the kinematic signatures e.g., (outflowing material from star-forming
regions) have been performed \citep{Steidel1992,Nestor2005,Prochter2006}.
Based on 4000 Mg {\sc ii} absorbers from zCOSMOS \citep{Lilly2007},  
\citet{Bordoloi2011} find that at the same 
stellar mass, the strength of Mg {\sc ii} absorption is much higher for blue
star-forming galaxies than for red galaxies. 
  Determinations of the Mg abundances from the absorption lines in 
damped Ly$\alpha$ absorbing (DLA) systems gave Mg/O abundance
ratios that are close to the solar values. 
  Meanwhile, the absorption line systems allow investigations of gas properties
in outer parts of galaxies and in the intergalactic medium.

Element abundances of the interstellar
medium are usually compared with the solar abundances, assuming that the 
interstellar medium has the same composition as the Sun, which 
is, however, not the case \citep{Snow1996}.
 Instead, 
the depletion will be different if the reference standard is derived from 
different types of stars in the solar neighbourhood,  or the
solar neighbourhood abundances corrected for the chemical evolution 
of the Galaxy, since formation of the Sun, 
or taking the galactocentric gradient of the chemical elements
into account \citep{Snow1996,PP10}.
 Thus, \citet{Snow1996} found systematically lower depletions using different
so-called ``cosmic''  abundances as reference.
   Different authors also use different solar abundances 
as reference abundance \citep[e.g.][]{Grevesse1998,Asplund2005,A09} or
solar system meteoritic  values, and different metallicity 
indicators (O, S, Zn) in DLA, QSO, and GRB absorption systems.
  Using such different reference abundances resulted in a Mg 
depletion that does
not exceed $\sim$0.2 dex. \citep[Throughout the paper we use 
solar values by ][]{A09}. 

  However, despite numerous magnesium abundance determinations, 
only upper limits of Mg {\sc ii} abundances have been obtained in the 
overwhelming majority of the cases. 
There is also a problem with some O~{\sc i} absorption oxygen lines 
since those are generally saturated. Therefore
there are not so many exact Mg/O abundance ratios derived in DLAs.

 No such determinations exist for extragalactic H~{\sc ii} regions,
excluding the Tarantula nebula.
This is because of the weakness of the 
Mg~{\sc ii} $\lambda$4481\AA\ recombination line seen in emission,
and its blending with other weak emission lines.
The only possibility for studing the Mg abundance in
extragalactic H~{\sc ii} regions
is to use intensities of the resonance doublet 
Mg~{\sc ii} $\lambda$2797, $\lambda$2803 lines, which are much
brighter than the recombination lines. 
 However, these 
emission lines, at variance to forbidden lines, are subject to absorption 
by the weakly ionised interstellar medium. 
  It is also possible that in some galaxies gas outflows can be
responsible for an additional underlying contribution to the Mg~{\sc ii} lines
producing broad features due to the Doppler effect.
However, for our galaxy sample we cannot consider the effects of Mg gas 
outflows and radiation
transfer in the medium surrounding the H~{\sc ii} regions 
due to the unknown gas velocity distribution and ionisation structure of 
our objects.
 Additionally, Mg~{\sc ii} $\lambda$2797, $\lambda$2803 emission and absorption
can be produced by cool giant and supergiant stars, and luminous blue
variable (LBV) stars. 

Finally, the doublet in spectra of low-redshift galaxies
can only be observed from space. However, in higher redshift galaxies,
Mg~{\sc ii} $\lambda$2797, $\lambda$2803 lines can be measured in the 
optical range. 
  In particular, these lines are seen in the SDSS spectra 
with a lower wavelength cut-off of $\sim$3800\AA\
if the galaxy redshift $z$ is greater than $\sim$ 0.36, allowing the 
determination of its abundance.

The aim of this paper is to select SDSS spectra of emission-line star-forming
galaxies with $z$ $\ga$ 0.36 showing the 
Mg~{\sc ii} $\lambda$2797, $\lambda$2803 doublet in emission and to derive Mg
abundances.
  Our sample of emission-line galaxies extracted from the SDSS is discussed 
in Section \ref{S2}. The element abundances are
derived in Section \ref{S3}. We discuss our results and compare them with
other types of objects in Section \ref{S4}. 
Our main findings are summarised in Section \ref{S5}.

\setcounter{table}{0}

\begin{table*}
 \caption{General characteristics of galaxies}
 \label{tab1}
\begin{tabular}{ccccccccccc} \hline
Name        & R.A.(J2000)&Dec.(J2000)      & $z$    &$g$$^a$&&Name        & R.A.(J2000)&Dec.(J2000)      & $z$    &$g$$^a$\\ \hline
J0006$-$0903& 00:06:14.94&   $-$09:03:16.59&  0.4296& 19.61 &&J0925$+$2709& 09:25:49.14&   $+$27:09:28.45&  0.4891& 20.03 \\
J0101$-$0057& 01:01:05.92&   $-$00:57:29.65&  0.3903& 20.39 &&J0956$+$3203& 09:56:21.73&   $+$32:03:54.89&  0.4263& 20.62 \\
J0104$+$0013& 01:04:45.44&   $+$00:13:25.28&  0.4909& 20.38 &&J0957$+$3314& 09:57:24.75&   $+$33:14:08.14&  0.5300& 21.17 \\
J0142$+$0108& 01:42:57.53&   $+$01:08:24.47&  0.5242& 19.79 &&J1000$+$3324& 10:00:59.49&   $+$33:24:17.95&  0.3890& 21.61 \\
J0149$+$0100& 01:49:47.60&   $+$01:00:09.70&  0.5660& 20.65 &&J1002$+$3228& 10:02:21.48&   $+$32:28:26.38&  0.4854& 21.66 \\
J0155$-$0107& 01:55:28.88&   $-$01:07:47.01&  0.4146& 20.64 &&J1007$+$0446& 10:07:58.98&   $+$04:46:26.24&  0.4167& 20.44 \\
J0207$+$0047& 02:07:48.00&   $+$00:47:35.43&  0.5400& 19.98 &&J1045$+$3225& 10:45:29.29&   $+$32:25:32.13&  0.4094& 19.42 \\
J0232$-$0021& 02:32:41.75&   $-$00:21:23.53&  0.3678& 19.82 &&J1048$+$6359& 10:48:39.03&   $+$63:59:21.55&  0.3738& 20.21 \\
J0301$-$0806& 03:01:39.58&   $-$08:06:45.20&  0.4314& 21.84 &&J1056$+$4758& 10:56:42.43&   $+$47:58:50.81&  0.4347& 21.45 \\
J0319$-$0104& 03:19:59.81&   $-$01:04:49.98&  0.6295& 20.30 &&J1108$+$6344& 11:08:54.82&   $+$63:44:05.59&  0.4395& 20.51 \\
J0340$-$0048& 03:40:47.92&   $-$00:48:48.12&  0.4503& 22.11 &&J1112$+$6331& 11:12:19.22&   $+$63:31:48.85&  0.4115& 21.08 \\
J0747$+$3147& 07:47:54.23&   $+$31:47:16.59&  0.3781& 21.59 &&J1128$-$0317& 11:28:47.68&   $-$03:17:17.48&  0.5563& 20.34 \\
J0749$+$3142& 07:49:48.00&   $+$31:42:49.11&  0.6959& 21.11 &&J1135$+$6025& 11:35:27.96&   $+$60:25:32.99&  0.4299& 19.13 \\
J0756$+$3057& 07:56:43.78&   $+$30:57:29.49&  0.4213& 21.68 &&J1136$-$0223& 11:36:27.54&   $-$02:23:17.80&  0.3931& 20.92 \\
J0757$+$3148& 07:57:27.88&   $+$31:48:59.87&  0.6148& 21.43 &&J1213$+$2705& 12:13:09.77&   $+$27:05:38.58&  0.3830& 19.35 \\
J0804$+$1748& 08:04:23.50&   $+$17:48:25.17&  0.5498& 21.27 &&J1408$+$0224& 14:08:06.07&   $+$02:24:09.19&  0.4041& 20.22 \\
J0807$+$4002& 08:07:59.23&   $+$40:02:18.74&  0.4712& 21.15 &&J1420$+$2628& 14:20:01.22&   $+$26:28:04.29&  0.3872& 19.34 \\
J0812$+$3200& 08:12:41.86&   $+$32:00:46.38&  0.4562& 21.46 &&J1430$+$4802& 14:30:55.90&   $+$48:02:01.58&  0.4775& 20.70 \\
J0813$+$3339& 08:13:36.62&   $+$30:39:35.60&  0.5584& 20.93 &&J1544$+$3308& 15:44:18.78&   $+$33:08:47.84&  0.4000& 19.80 \\
J0815$+$3129& 08:15:24.40&   $+$31:29:36.05&  0.4055& 20.83 &&J1548$+$0727& 15:48:14.67&   $+$07:27:10.13&  0.4170& 20.20 \\
J0817$+$3159& 08:17:06.28&   $+$31:59:00.75&  0.4083& 20.79 &&J1555$+$3543& 15:55:16.39&   $+$35:43:24.65&  0.4519& 19.79 \\
J0837$+$3108& 08:37:57.41&   $+$31:08:40.39&  0.5144& 21.00 &&J1559$+$0634& 15:59:08.55&   $+$06:34:38.86&  0.5369& 20.00 \\
J0838$+$3057& 08:38:57.85&   $+$30:57:50.22&  0.5552& 21.74 &&J1602$+$4002& 16:02:06.36&   $+$40:02:49.11&  0.3885& 00.00 \\
J0844$+$3241& 08:44:31.93&   $+$32:41:05.99&  0.5988& 21.75 &&J1616$+$2057& 16:16:42.64&   $+$20:57:20.11&  0.5988& 20.30 \\
J0855$+$0318& 08:55:58.30&   $+$03:18:07.47&  0.4359& 21.66 &&J1716$+$2744& 17:16:00.88&   $+$27:44:14.40&  0.4867& 19.50 \\
J0904$+$2806& 09:04:46.04&   $+$28:06:05.65&  0.3859& 19.66 &&J2145$+$0040& 21:45:51.41&   $+$00:40:30.93&  0.4179& 20.38 \\
J0914$+$4451& 09:14:57.59&   $+$44:51:10.46&  0.4430& 21.12 &&J2208$-$0106& 22:08:40.38&   $-$01:06:12.29&  0.5115& 20.78 \\
J0915$+$3344& 09:15:23.10&   $+$33:44:29.77&  0.3948& 21.51 &&J2211$+$0114& 22:11:45.46&   $+$01:14:08.44&  0.3721& 20.35 \\
J0918$+$3301& 09:18:10.90&   $+$33:01:42.09&  0.5112& 21.50 &&J2334$-$0046& 23:34:36.23&   $-$00:46:36.61&  0.5245& 20.34 \\
J0918$+$3331& 09:18:28.35&   $+$33:31:06.56&  0.6334& 21.37 &&J2348$-$0041& 23:48:09.07&   $-$00:41:01.66&  0.4095& 20.22 \\
J0922$+$3330& 09:22:38.44&   $+$33:30:13.29&  0.4267& 21.82 &&J2352$+$0025& 23:52:37.96&   $+$00:25:58.77&  0.6354& 20.17 \\ \hline
\end{tabular}
$^a$Apparent SDSS $g$ magnitude.
\end{table*}

\begin{figure}
\hspace*{0.0cm}\psfig{figure=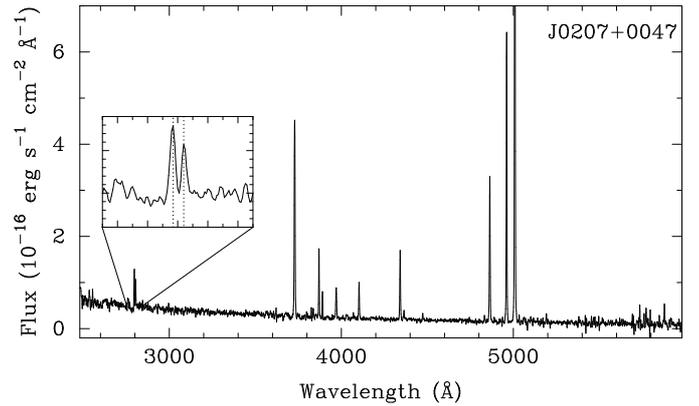,angle=-90,width=8.8cm,clip=}
\caption{The redshift-corrected SDSS spectrum of J0207$+$0047. The spectral
region $\lambda$$\lambda$2750 - 2850, which includes the
Mg~{\sc ii} $\lambda$2797, $\lambda$2803 emission lines, is shown in the inset.
Vertical dotted lines in the inset indicate the nominal wavelengths of the 
lines.}
\label{fig1}
\end{figure}

\begin{figure*}
\centering{
\hspace*{0.0cm}\psfig{figure=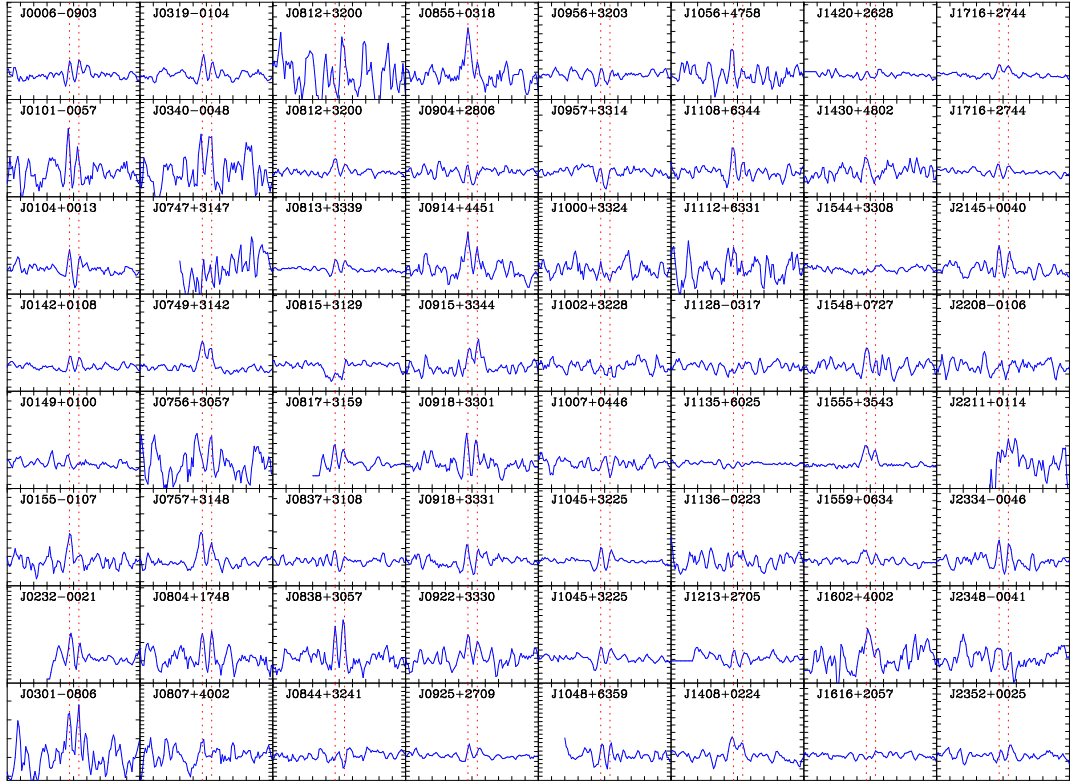,angle=-90,width=14.cm,clip=}
}
\caption{The redshift-corrected SDSS spectra in the wavelength range
$\lambda$$\lambda$2750 - 2850\AA\ of all selected galaxies excluding
J0207$+$0047, which is shown in Fig. \ref{fig1}. Vertical dotted lines
indicate the nominal wavelengths of the 
Mg~{\sc ii} $\lambda$2797, $\lambda$2803 
emission lines. For three galaxies, J0812+3200, J1045+3225, and J1716+2744, 
two spectra are shown that available in the SDSS data base.}
\label{fig2}
\end{figure*}

\section {Sample of SDSS galaxies \label{S2}}

We use a sample that is composed of spectra of low-metallicity 
H~{\sc ii} regions with strong emission lines selected from the SDSS DR7. 
The SDSS \citep{Y00} offers a gigantic
data base of galaxies with well-defined selection criteria that are observed 
in a homogeneous way. First, we extracted 
$\sim$ 15000 spectra with strong emission lines from the whole data base of
$\sim$ 800000 galaxy spectra. Out of this sample we selected 65 spectra
of star-forming galaxies with strong nebular H$\beta$, 
H$\alpha$, [O~{\sc ii}]$\lambda$3727, 
[O~{\sc iii}]$\lambda$4959, $\lambda$5007 emission lines and redshifts 
$z$ $\ga$ 0.36. 
  The [O~{\sc iii}]$\lambda$4363 emission line is also present in many of 
these spectra (in $\sim$63\% of the total sample and $\sim$70\% 
of the galaxies in which Mg~{\sc ii} lines were detected), allowing an 
accurate determination of the
oxygen abundance by the direct $T_e$-method. For three galaxies, 
two spectra are available in the SDSS data base. Therefore, the total 
number of selected galaxies is 62.

The general characteristics of the selected galaxies are shown in 
Table \ref{tab1}. The redshift range is 0.36 -- 0.70. 
In general these galaxies are very faint, with
the apparent SDSS $g$ magnitude fainter than 19 mag.
 The redshift-corrected spectrum of one galaxy, J0207+0047, is shown in 
Fig. \ref{fig1}. This spectrum resembles the spectrum of high-excitation 
star-forming H~{\sc ii} region with strong emission lines, including
[O~{\sc iii}]$\lambda$4363. The 
Mg~{\sc ii} $\lambda$2797, $\lambda$2803 doublet is also present in
emission and it is shown in more detail in the inset. The remaining
64 redshift-corrected spectra in the wavelength range 
$\lambda$$\lambda$2750 -- 2850\AA\ are shown in Fig. \ref{fig2}. 
Despite the noisy spectra in some cases, the Mg~{\sc ii} emission is detected in
about two thirds of the spectra. However, 
we also note that blue-shifted Mg~{\sc ii} absorption is present in many 
cases, which may be an indication of stellar
winds from cool massive stars, such as LBVs.
  In some cases broad absorption
features are observed at the position of 
Mg {\sc ii} $\lambda$2797\AA\ and $\lambda$2803\AA\ emission lines. 
The most prominent feature of this kind is seen in the spectrum of J0815+3129.
In some other cases absorption features are absent in the spectra.
  We take all these peculiarities into account  
and measure the emission line intensities by placing a continuum level
at the bottom of the absorption profiles if they are present.

\setcounter{table}{2}

\begin{table}
\caption{Input parameters for the grid of the photoionised H~{\sc ii}
region models \label{tab3}}
\begin{tabular}{lc} \hline \hline
Parameter                  & Value   \\ \hline
log $Q$(H)$^a$             &  52, 53, 54 \\
Starburst age, Myr         &  2.0, 3.5, 4.0    \\
$N_e$$^b$                  &  10, 10$^2$, 10$^3$, var \\
log $f$$^c$                & $-$0.5, $-$1.0, $-$1.5, $-$2.0 \\
Oxygen abundance 12+logO/H & 7.3, 7.6, 8.0, 8.3 \\
$^4$He mass fraction $Y$       &  0.254   \\ \hline
\end{tabular}

$^a$log of the number of ionising photons in units s$^{-1}$.

$^b$The electron number density in units cm$^{-3}$. The electron
number density in models labelled ``var'' is varied with
radial distance according to Eq.~\ref{eq1}.

$^c$log of volume filling factor.

\end{table}

\begin{table}
 \caption{Coefficients for the $ICF$(Mg$^+$)$^a$ fit in Eq. \ref{eq5}}
 \label{tab4}
\begin{tabular}{cccc} \hline
$i$        & $a_i$                  &   $b_i$                  &$c_i$  \\ \hline
1          &4.4102$\times$10$^{-21}$&   2.2063$\times$10$^{2}$ &5.3269$\times$10$^{1}$ \\
2          &9.2000$\times$10$^{-9}$ &   2.5726$\times$10$^{1}$ &2.3156$\times$10$^{1}$ \\
3          &9.2671$\times$10$^{-4}$ &$-$2.2616$\times$10$^{0}$ &1.0263$\times$10$^{1}$ \\
4          &1.3081$\times$10$^{0}$  &$-$2.7979$\times$10$^{-1}$&1.3332$\times$10$^{0}$ \\ \hline
\end{tabular}

$^a$$ICF$ is the ionisation correction factor.

\end{table}

\begin{figure}
\hspace*{0.0cm}\psfig{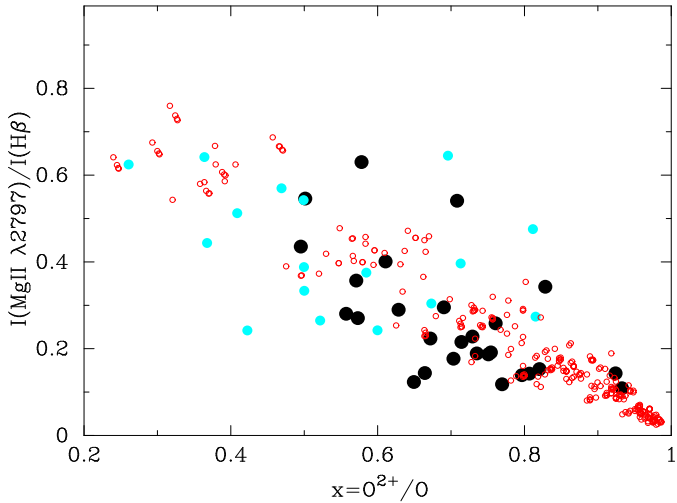}
\caption{Dependence of $I$(Mg~{\sc ii} $\lambda$2797)/$I$(H$\beta$) on the
excitation parameter $x$ = O$^{2+}$/O. Red open circles are data
from CLOUDY models calculated by adopting 12 + logO/H = 8.0 and 8.3.
 {\textbf Large black filled circles  are extinction-corrected intensities for
the H {\sc ii} regions with
12 + logO/H $\geq$ 7.9 and small light blue filled circles 
with 12 + logO/H $<$ 7.9.} }
\label{fig3}
\end{figure}

   The extinction-corrected line fluxes $I$($\lambda$), normalised to 
$I$(H$\beta$), are given in Table \ref{tab2}. They are only
available in the electronic online version. 
The line fluxes were obtained using the IRAF\footnote {IRAF is the Image 
Reduction and Analysis Facility distributed by the National Optical Astronomy 
Observatory, which is operated by the Association of Universities for Research 
in Astronomy (AURA) under cooperative agreement with the National Science 
Foundation (NSF).} SPLOT routine.
The line flux errors 
include statistical errors 
in addition to errors introduced by
the standard star absolute flux calibration, which we set to 1\% of the
line fluxes. 
  These errors will be propagated later into the calculation
of abundance errors.
The line fluxes were corrected for two effects: (1) reddening using 
the extinction curve of \citet{C89} 
and (2) underlying hydrogen stellar absorption derived simultaneously by an 
iterative procedure as described in \citet{ITL94}.
  Since the redshifts of the selected galaxies are high, the correction for
extinction was done in two steps. First, emission-line intensities with
observed wavelengths were corrected for the Milky Way extinction, 
using values of the extinction  $A$($V$) in the $V$ band from the 
NASA/IPAC extragalactic database (NED).
Then, the internal extinction was derived from the Balmer hydrogen emission 
lines, corrected for the Milky Way extinction. The internal extinction was
applied to correct line intensities at non-redshifted wavelengths.
The extinction coefficients in both cases of the Milky Way and the internal 
extinction are defined as $C$(H$\beta$) = 1.47$E(B-V)$,
where $E(B-V)$ = $A(V)$/3.2 \citep{A84}. 
  The mean value of the internal
extinction coefficient  $C$(H$\beta$) $\sim$ 0.2 is typical
of star-forming galaxies. 

\setcounter{table}{4}

 \begin{table*}
 \caption{Spectroscopic parameters}
 \label{tab6}
 \begin{tabular}{lcccclccc} \hline
 Object & 12+logO/H&$[$Mg/O$]$&EW(H$\beta$)$^b$&&Object &   12+logO/H&$[$Mg/O$]$&EW(H$\beta$)$^b$ \\ \hline
J0006$-$0903 &7.96$\pm$0.05&$-$0.19$\pm$0.15&  50&&J0956$+$3203 &7.66$\pm$0.06&$+$0.00$\pm$0.00&  43\\
J0101$-$0057 &7.96$\pm$0.02&$-$0.03$\pm$0.10&  64&&J0957$+$3314 &7.78$\pm$0.02&$-$0.36$\pm$0.10&  39\\
J0104$+$0013 &8.12$\pm$0.06&$-$0.12$\pm$0.18&  49&&J1000$+$3324 &7.82$\pm$0.02&$+$0.00$\pm$0.00&  48\\
J0142$+$0108 &8.05$\pm$0.04&$-$0.31$\pm$0.14&  68&&J1002$+$3228 &7.58$\pm$0.02&$+$0.00$\pm$0.00&  50\\
J0149$+$0100 &7.96$\pm$0.02&...&  94&&J1007$+$0446 &7.80$\pm$0.01&...&  65\\
J0155$-$0107 &8.04$\pm$0.02&$-$0.49$\pm$0.11& 185&&J1045$+$3225$^a$&8.19$\pm$0.03&$-$0.45$\pm$0.10&  77\\
J0207$+$0047 &8.22$\pm$0.02&$-$0.24$\pm$0.09& 107&&J1045$+$3225$^a$&8.24$\pm$0.03&$-$0.38$\pm$0.11&  75\\
J0232$-$0021 &8.20$\pm$0.05&$-$0.58$\pm$0.17&  82&&J1048$+$6359 &7.57$\pm$0.01&$-$0.34$\pm$0.06&  42\\
J0301$-$0806 &7.83$\pm$0.01&$-$0.15$\pm$0.09& 167&&J1056$+$4758 &8.00$\pm$0.01&$-$0.24$\pm$0.11& 431\\
J0319$-$0104 &8.06$\pm$0.02&$-$0.36$\pm$0.08&  94&&J1108$+$6344 &8.04$\pm$0.02&$-$0.49$\pm$0.10& 108\\
J0340$-$0048 &7.89$\pm$0.01&$-$0.24$\pm$0.07& 199&&J1112$+$6331 &7.77$\pm$0.01&...& 132\\
J0747$+$3147 &7.84$\pm$0.03&...&  78&&J1128$-$0317 &7.95$\pm$0.02&...&  32\\
J0749$+$3142 &7.87$\pm$0.02&$+$0.00$\pm$0.09& 147&&J1135$+$6025 &7.66$\pm$0.05&...&  38\\
J0756$+$3057 &7.66$\pm$0.02&$-$0.13$\pm$0.07&  50&&J1136$-$0223 &7.96$\pm$0.01&...&  38\\
J0757$+$3148 &8.05$\pm$0.02&$-$0.04$\pm$0.09& 147&&J1213$+$2705 &7.72$\pm$0.01&$-$0.44$\pm$0.05&  35\\
J0804$+$1748 &8.06$\pm$0.02&$-$0.38$\pm$0.09& 165&&J1408$+$0224 &7.82$\pm$0.01&$-$0.14$\pm$0.08& 131\\
J0807$+$4002 &7.87$\pm$0.02&...& 120&&J1420$+$2628 &8.29$\pm$0.06&...&  59\\
J0812$+$3200$^a$&7.94$\pm$0.02&...& 339&&J1430$+$4802 &7.87$\pm$0.01&...&  45\\
J0812$+$3200$^a$&8.15$\pm$0.03&$-$0.44$\pm$0.17& 116&&J1544$+$3308 &7.62$\pm$0.01&...&  32\\
J0813$+$3339 &7.69$\pm$0.02&$-$0.20$\pm$0.11&  43&&J1548$+$0727 &8.19$\pm$0.03&$-$0.39$\pm$0.10& 126\\
J0815$+$3129 &7.64$\pm$0.02&...&  25&&J1555$+$3543 &8.31$\pm$0.05&$-$0.32$\pm$0.16&  95\\
J0817$+$3159 &8.25$\pm$0.04&$-$0.36$\pm$0.15& 117&&J1559$+$0634 &7.82$\pm$0.01&$-$0.34$\pm$0.08&  32\\
J0837$+$3108 &7.86$\pm$0.05&$-$0.58$\pm$0.21&  63&&J1602$+$4002 &7.94$\pm$0.01&...&  47\\
J0838$+$3057 &8.07$\pm$0.01&$-$0.53$\pm$0.12& 113&&J1616$+$2057 &8.03$\pm$0.04&...&  91\\
J0844$+$3241 &7.83$\pm$0.08&$-$0.15$\pm$0.25&  62&&J1716$+$2744$^a$&7.88$\pm$0.01&$-$0.34$\pm$0.04&  44\\
J0855$+$0318 &7.83$\pm$0.01&$-$0.04$\pm$0.06& 388&&J1716$+$2744$^a$&7.95$\pm$0.01&$-$0.37$\pm$0.04&  44\\
J0904$+$2806 &7.53$\pm$0.10&$-$0.20$\pm$0.26&  31&&J2145$+$0040 &7.84$\pm$0.01&$-$0.25$\pm$0.05&  66\\
J0914$+$4451 &7.92$\pm$0.01&$-$0.56$\pm$0.07& 170&&J2208$-$0106 &7.89$\pm$0.01&...&  71\\
J0915$+$3344 &8.11$\pm$0.09&$-$0.25$\pm$0.29& 207&&J2211$+$0114 &7.77$\pm$0.01&$-$0.53$\pm$0.09&  28\\
J0918$+$3301 &8.44$\pm$0.09&$-$0.02$\pm$0.30& 424&&J2334$-$0046 &7.91$\pm$0.01&$-$0.25$\pm$0.05&  54\\
J0918$+$3331 &7.52$\pm$0.11&$-$0.15$\pm$0.31&  61&&J2348$-$0041 &8.05$\pm$0.03&...& 100\\
J0922$+$3330 &8.00$\pm$0.02&$-$0.49$\pm$0.17& 131&&J2352$+$0025 &8.18$\pm$0.05&$-$0.41$\pm$0.18&  64\\
J0925$+$2709 &8.36$\pm$0.08&$-$0.56$\pm$0.28&  73\\
 \hline
 \end{tabular}
   
 $^a$Galaxies with two available spectra in the SDSS.

 $^b$in \AA.
 \end{table*}

Additionally, the Mg~{\sc ii} $\lambda$2797, $\lambda$2803 emission lines
were corrected for the underlying stellar absorption. For this, we
used the \citet{BC03} population synthesis models of single stellar populations.
Spectra of these models do not have sufficient spectral resolution to separate
Mg~{\sc ii} $\lambda$2797 and $\lambda$2803 absorption lines. 
Therefore, we measured the equivalent width of the blend. We find that in
a wide range of starburst ages of 3 -- 10 Myr, the equivalent width
EW$_{abs}$(Mg~{\sc ii} $\lambda$2797 + $\lambda$2803) is constant and is
equal to $\sim$ --1\AA. The ``--'' sign means that the line is in absorption.
Then, for separate Mg~{\sc ii} absorption lines we adopt equal
EW$_{abs}$'s of $-0.5$\AA\ and correct Mg~{\sc ii} emission lines
multiplying their intensities by a factor (EW+$|$EW$_{abs}$$|$)/EW, where
EW is the equivalent width of the emission line.
   Equivalent widths EW(H$\beta$), extinction coefficients 
$C$(H$\beta$)(MW) and 
$C$(H$\beta$)(int), and EW$_{abs}$ of the hydrogen
absorption stellar lines are also given in Table \ref{tab2}, 
along with the uncorrected H$\beta$ fluxes.

\begin{figure*}
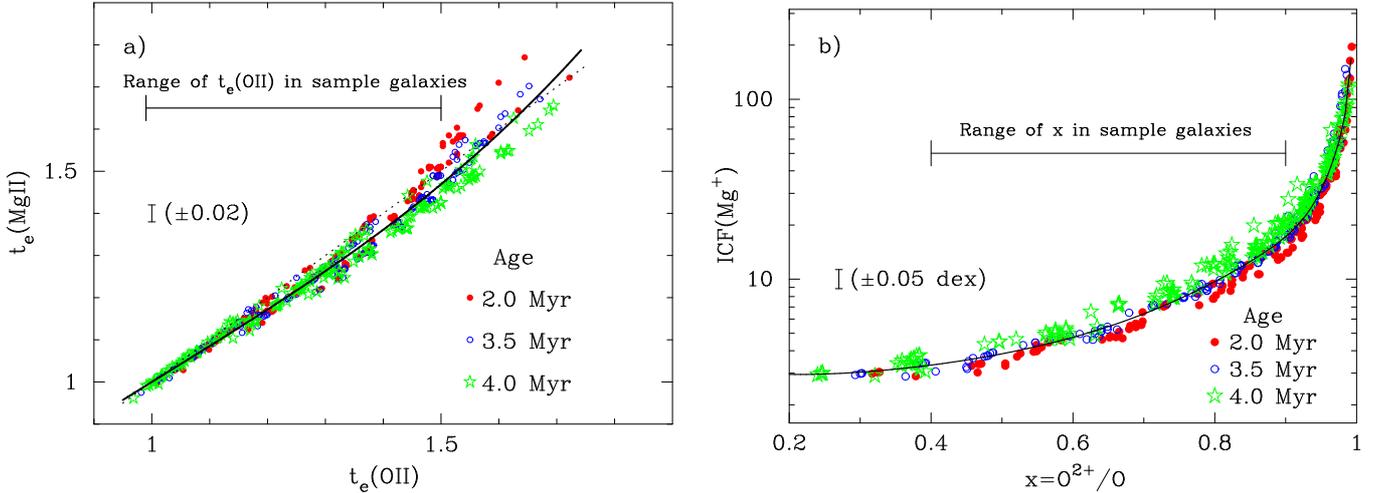

\hspace*{0.0cm}\psfig{figure=aa21010-12_fig4a.ps,angle=-90,width=8.8cm,clip=}
\hspace*{0.4cm}\psfig{figure=aa21010-12_fig4b.ps,angle=-90,width=8.56cm,clip=}
\caption{(a) The relation between the electron temperatures
$t_e$(Mg~{\sc ii}) and $t_e$(O~{\sc ii}). $t_e$ are in units of
10$^{-4}$$T_e$. Red filled circles, blue open circles, and green stars are data 
from CLOUDY models with starburst ages of 2.0 Myr, 3.5 Myr, and 4.0 Myr,
respectively. The solid line is the fit to data with all ages, defined by 
Eq. \ref{eq2}. The dotted
line is the line of equal temperatures. The range of the electron temperature
$t_e$(O~{\sc ii}) in the sample galaxies is shown by a horizontal line.
The error bar is the dispersion of $t_e$(Mg~{\sc ii}) in the range of
$t_e$(O~{\sc ii}) shown by the horizontal line.
  (b) The relation between the
ionisation correction factor $ICF$(Mg$^+$) and the excitation parameter
$x$ = O$^{2+}$/O. Red filled circles, blue open circles, and green stars are 
data from CLOUDY models with starburst ages of 2.0 Myr, 3.5 Myr, and 4.0 Myr,
respectively. The solid line is the fit to data with all ages as defined by 
Eq. \ref{eq5} with
the coefficients from Table \ref{tab3}. The range of the excitation parameter
$x$ in the sample galaxies is shown by a horizontal line.
The error bar is the dispersion of $ICF$(Mg$^+$) (in dex) in the range of
$x$ shown by the horizontal line.
}
\label{fig4}
\end{figure*}

\section {Element abundances \label{S3}}

\subsection{Oxygen abundance} \label{O}

   To determine element abundances, we generally follow  
the procedures of \citet{ITL94,ITL97} and \citet{TIL95}.
We adopt a two-zone photoionised H {\sc ii}
region model: a high-ionisation zone with temperature $T_e$(O~{\sc iii}), 
where the [O~{\sc iii}] lines
originate, and a 
low-ionisation zone with temperature $T_e$(O~{\sc ii}), where the [O~{\sc ii}]
lines originate. 
In the H {\sc ii} regions with a detected [O~{\sc iii}] $\lambda$4363
emission line, the temperature $T_e$(O~{\sc iii}) is calculated using the 
direct method based on the 
[O~{\sc iii}] $\lambda$4363/($\lambda$4959+$\lambda$5007) line ratio.
In H {\sc ii} regions where the [O~{\sc iii}] $\lambda$4363 emission line 
is not detected, we used a semi-empirical method described by \citet{IT07} to 
derive $T_e$(O~{\sc iii}).
For $T_e$(O~{\sc ii}), we use
the relation between the electron temperatures $T_e$(O~{\sc iii}) and
$T_e$(O~{\sc ii}) obtained by \citet{I06} from
the H {\sc ii} photoionisation models \citep{SI03}. 

  Ionic and total oxygen abundances are derived
using expressions for ionic abundances  
obtained by \citet{I06}.
For magnesium, we assume that the Mg~{\sc ii} emission has a
nebular origin (see text below and Fig.~\ref{fig3}). Then the 
magnesium abundance can be derived if the electron 
temperature $T_e$(Mg~{\sc ii}) in the Mg$^+$ zone and the ionisation 
correction factor $ICF$(Mg$^+$) for unseen stages of ionisation are known.
  To derive $T_e$(Mg~{\sc ii}) and $ICF$(Mg$^+$) we use a grid of the 
photoionised H~{\sc ii} region models that also predict 
Mg~{\sc ii} $\lambda$2797, $\lambda$2803 emission-line intensities.

\begin{figure*}
\hspace*{0.0cm}\psfig{figure=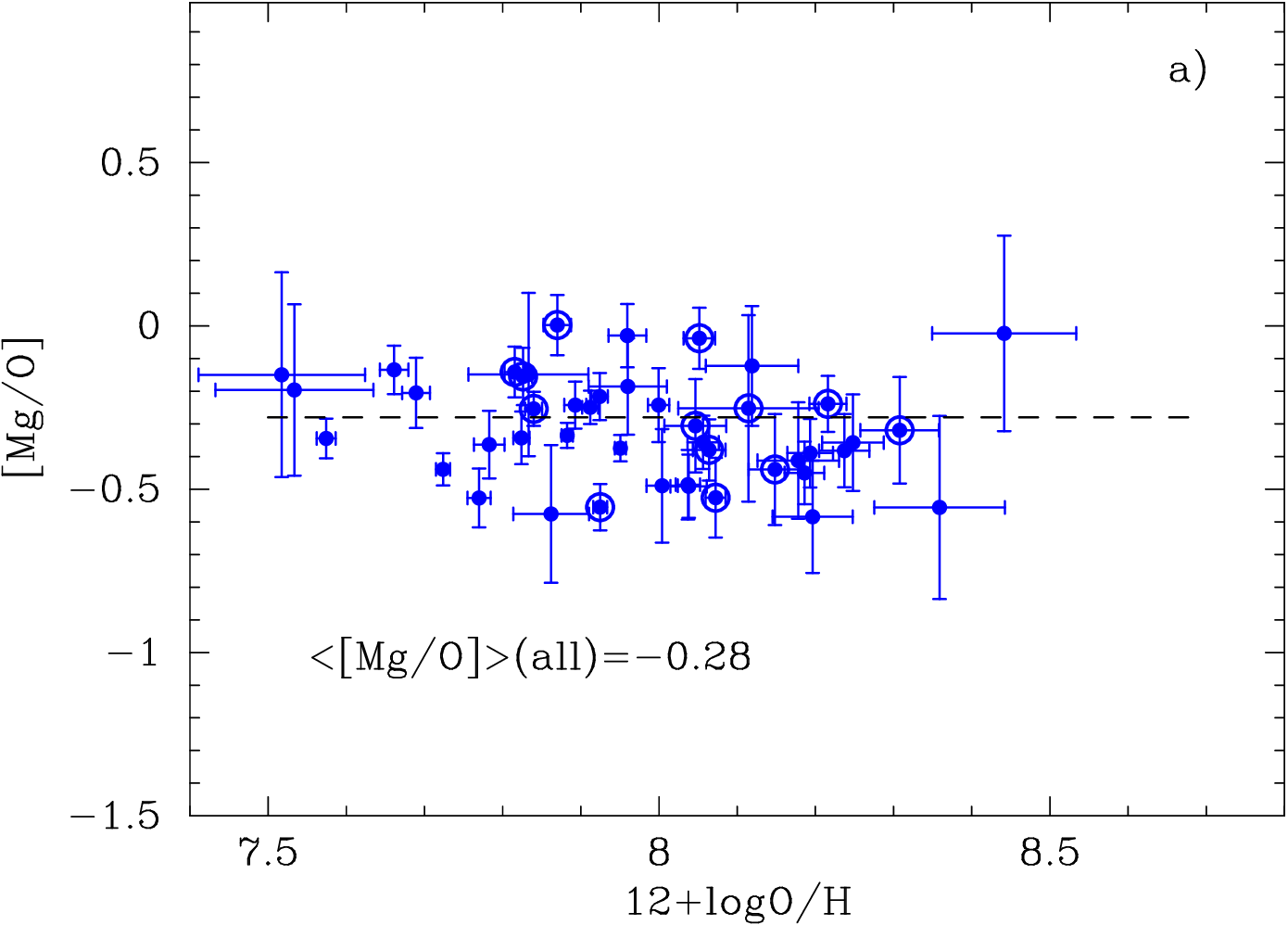,angle=-90,width=8.8cm,clip=}
\hspace*{0.4cm}\psfig{figure=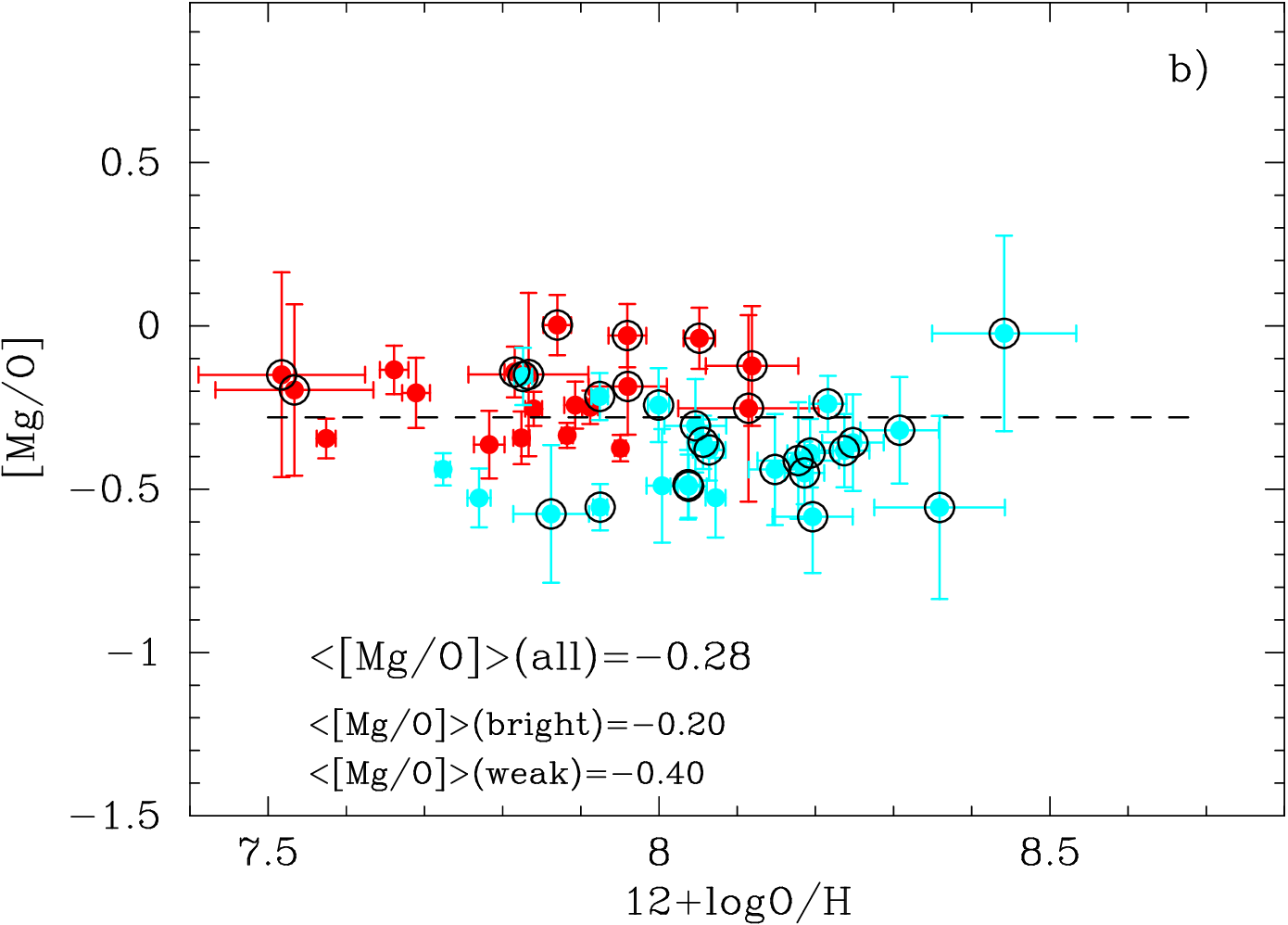,angle=-90,width=8.8cm,clip=}
\caption{Dependence of the magnesium-to-oxygen abundance ratio relative
to the solar value, [Mg/O] = log(Mg/O) -- log(Mg/O)$_\odot$, on oxygen
abundance 12+logO/H with error bars for all data points.  
The solar value log(Mg/O)$_\odot$ = --1.09 
is adopted from \citet{A09}.
Dashed line and $<$[Mg/O]$>$ in both panels are mean values of [Mg/O].
(a) The encircled galaxies are those with no clear presence of blue-shifted 
absorption profiles (Fig. \ref{fig2}).
(b) The same sample as in (a) but with division between bright
Mg {\sc ii} lines with $I$($\lambda$2797\AA)/$I$(H$\beta$) $\geq$ 0.3 
(red filled circles) and weak Mg {\sc ii} lines 
with $I$($\lambda$2797\AA)/$I$(H$\beta$) $<$ 0.3 (blue filled circles).
  Encircled galaxies in (b) are those where 
[O~{\sc iii}]$\lambda$4363\AA\ is measured.
Additionally, the mean values of [Mg/O] for bright ($<$[Mg/O]$>$(bright))
and weak ($<$[Mg/O]$>$(weak)) Mg {\sc ii} lines are denoted.
}
\label{fig5}
\end{figure*}

\subsection{Grid of photoionisation CLOUDY models}\label{cloudy}

Using the version v10.00 of the CLOUDY code \citep{F98} we calculated a grid
of 432 spherical ionisation-bounded H~{\sc ii} region models with parameters 
shown in Table \ref{tab3},
which cover the entire range of parameters in real high-excitation,
low-metallicity H~{\sc ii} regions.
In particular, the range of oxygen abundances is 
12 + log O/H $\approx$ 7.5 -- 8.4 for our sample.
 The abundances of other heavy elements relative to oxygen are kept 
constant and correspond to the typical value obtained for low-metallicity
emission-line galaxies \citep[e.g. ][]{I06}. We also include dust, scaling it
according to the oxygen abundance. The characteristics adopted for the dust 
are those offered by CLOUDY as ``Orion nebula dust''. 

We adopt three values of the number of ionising photons Q and the shape
for the ionising radiation spectrum corresponding to the Starburst99
model with the ages of 2.0, 3.5, and 4.0 Myr and different metallicities 
\citep{L99}. 
Thus, for 12+logO/H = 7.3 and 7.6 we adopt Starburst99
models with the heavy element mass fraction $Z$ = 0.001, for those
with 12+logO/H = 8.0 models with $Z$ = 0.004 and for those 
with 12+logO/H = 8.3 models with $Z$ = 0.008.
All Starburst99 models were calculated with the \citet{HM98} and 
\citet{P01} stellar atmosphere set
and with the stellar tracks from \citet{M94}. 
We also vary the log of volume - filling factor $f$ between $-$0.5 
and $-$2.0 to obtain CLOUDY models with different ionisation parameters. 

Our range of the number density, $N_e$ = 10 -- 10$^3$ cm$^{-3}$, 
which is kept constant along a given
H~{\sc ii} region radius, covers the whole range expected for the 
extragalactic H~{\sc ii} regions.
Additionally, we calculated a set of 144 H~{\sc ii} region models with
parameters from Table \ref{tab3} (excluding $N_e$) with a Gaussian 
density distribution as a function of radius $r$ according to
\begin{equation}
N_e(r)=N_e(0)\exp\left[-\frac{r^2}{(30\,{\rm pc})^2}\right], \label{eq1}
\end{equation}
where $N_e(0)$ = 10$^3$ cm$^{-3}$. 
Thus, the total number of the models, which we use
for the subsequent analysis, is 576.

  Since the CLOUDY code allows us to predict Mg~{\sc ii} line intensities, we 
compared these intensities with the observed ones. 
   The result of the comparison is shown in Fig. \ref{fig3} 
where we show the 
Mg~{\sc ii} $\lambda$2797 line intensity 
as a function of the excitation parameter $x$=O$^{2+}$/O. 
  The observed H {\sc ii} regions with 12 + logO/H $\geq$ 7.9 
are shown by large black filled circles 
and H {\sc ii} regions with 12 + logO/H $<$ 7.9 by small light blue 
filled circles.
 Thus, the range of oxygen abundances in most of our galaxies is 
 12+logO/H $\sim$ 7.7 -- 8.4 (see Table \ref{tab6}). 
  Therefore, we show
in Fig. \ref{fig3} only predicted Mg~{\sc ii} $\lambda$2797 line intensities 
in the models with 12+logO/H = 8.0 and 8.3 (red open circles). 
   Our comparison shows that in general 
extinction-corrected Mg~{\sc ii} $\lambda$2797 observed intensities 
are by a factor of $\sim$1.3 lower than the predicted ones even for 
the H {\sc ii} regions with 12 + logO/H between 7.9 and 8.4, 
but they follow a trend with $x$ 
which is similar to the trend for predicted intensities. This implies that
Mg~{\sc ii} $\lambda$2797 emission is most likely nebular in origin, and 
the reduced line intensities are due to the combined effect of interstellar
absorption and Mg depletion onto dust.

\subsection{Magnesium abundance}\label{Mg}

The electron temperature $T_e$(Mg {\sc ii}) in the Mg$^+$ zone 
and $T_e$(O~{\sc ii}) in the O$^+$ zone are obtained 
from the CLOUDY photoionised H~{\sc ii} region models. In Fig. \ref{fig4}a
we show the relation between the electron temperatures in the O$^+$ and Mg$^+$
zones by different symbols for different starburst ages.
 As expected, these temperatures are very similar.  
 The dispersion of the data at high 
temperatures $T_e$(O~{\sc ii}) $\geq$ 15000K is mainly due to the different
starburst ages. However, at lower temperatures, which is the case for our
galaxies, the dispersion
is small and there is no evident separation between models with different
starburst ages. We fit the relation for models with all ages by an
expression
\begin{eqnarray}
t_e({\rm Mg~\textsc{ii}})&=&0.4560t_e({\rm O~\textsc{ii}})^3-1.4262t_e({\rm O~\textsc{ii}})^2 \nonumber \\
                         & &+2.3378t_e({\rm O~\textsc{ii}})-0.3675, \label{eq2}
\end{eqnarray}
which is shown in Fig. \ref{fig4}a by solid line. We will use this fit in
our subsequent analysis. However, we note that the fit in Eq. \ref{eq2}
is only applicable to the range $T_e$(O~{\sc ii}) $\sim$ 9000 -- 17000K.

The Mg$^+$ abundance is derived from the equation
\begin{eqnarray}
\frac{\rm Mg^+}{\rm H^+} &=& 
2.63 \times 10^{-8}t_e({\rm Mg~\textsc{ii}})^{-0.482}10^{2.186/t_e({\rm Mg~\textsc{ii}})} \nonumber \\
& & \times \frac{I({\rm Mg~\textsc{ii}~\lambda 2797+ \lambda 2803})}{I (\rm H\beta)}, \label{eq3}
\end{eqnarray}
where we use collisional strengths for the Mg~{\sc ii} $\lambda$2797 and 
$\lambda$2803 transitions from \citet{M83}. These 
collisional strengths are in very good agreement with those obtained
more recently by \citet{SP95}. 

The total Mg abundance is obtained from the relation
\begin{equation}
\frac{\rm Mg}{\rm H} = ICF({\rm Mg^+})\frac{\rm Mg^+}{\rm H^+}, \label{eq4}
\end{equation}
where the ionisation correction factor $ICF$(Mg$^+$) takes all unseen stages 
of magnesium ionisation inside the H~{\sc ii} region into account.
To derive $ICF$(Mg$^+$) we use our grid CLOUDY models and define
it as the ratio $x$(H$^+$)/$x$(Mg$^+$), where $x$(H$^+$) = H$^+$/H and 
$x$(Mg$^+$) = Mg$^+$/Mg
are volume-averaged fractions of H$^+$ and Mg$^+$, respectively.

In Fig. \ref{fig4}b we show the dependence of $ICF$(Mg$^+$) on the
excitation parameter O$^{2+}$/O. 
   Ionisation potentials of Mg of the three ionisation stages Mg, 
 Mg$^{+}$, and Mg$^{++}$ are 7.6, 15.0 and 80.1 eV, respectively. Therefore,
most of magnesium in a photoionised region is in the
Mg$^{+}$ and Mg$^{++}$ stages. However, because 
Mg$^{+}$ is not the dominant ionic species for most nebulae and
because Mg$^{++}$ is not observed in the visible range, a rather 
large $ICF$(Mg$^+$) is needed.
   Models with different starburst
ages in Fig. \ref{fig4}b are shown by different symbols. There is a clear 
offset between
the models with younger and older starburst ages, indicating differences
in the spectral energy distribution of the ionising radiation. 
  The fit to the data with all ages defined by equation 
\begin{equation}
ICF({\rm Mg^+}) = 
\sum_{i=1}^4 a_it^{b_i}_e \exp\left(\frac{c_i}{t_e}\right), \label{eq5}
\end{equation}
where coefficients $a_i$, $b_i$, and $c_i$ are given in Table \ref{tab4}.
   The fit is produced for a wide range of $x$ = 0.2 -- 1.0 where
$ICF$ increases from $\sim$ 3 to more than 100. However, the range
of $x$ in the sample galaxies (horizontal line) is smaller, 
corresponding to $ICF$s in the range $\sim$ 3.5 -- $\sim$ 20 with an average
value of $\sim$ 6. The fit closely follows the models with starburst age 
of 3.5 Myr corresponding to the equivalent width 
EW(H$\beta$) $\sim$ 100\AA\ that is typical for our galaxies.

\begin{figure}
\hspace*{0.2cm}\psfig{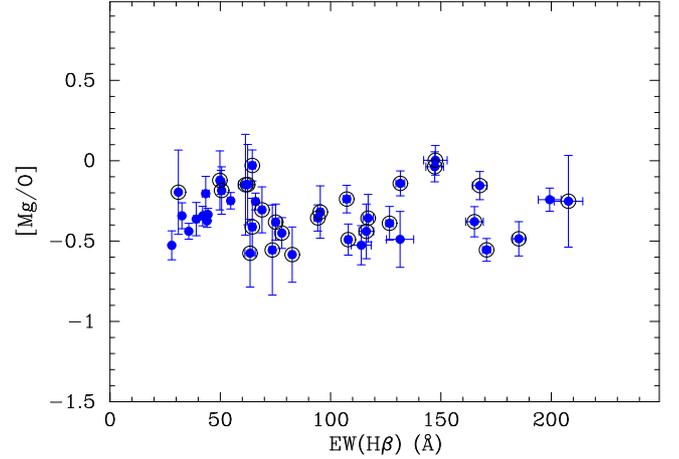}
\caption{Dependence of the magnesium-to-oxygen abundance ratio relative
to the solar value [Mg/O] 
on the equivalent width EW(H$\beta$) of the H$\beta$
emission line. 
   The galaxies with the measured  
[O~{\sc iii}]$\lambda$4363\AA\ line are encircled.
}
\label{fig6}
\end{figure}

\begin{figure}
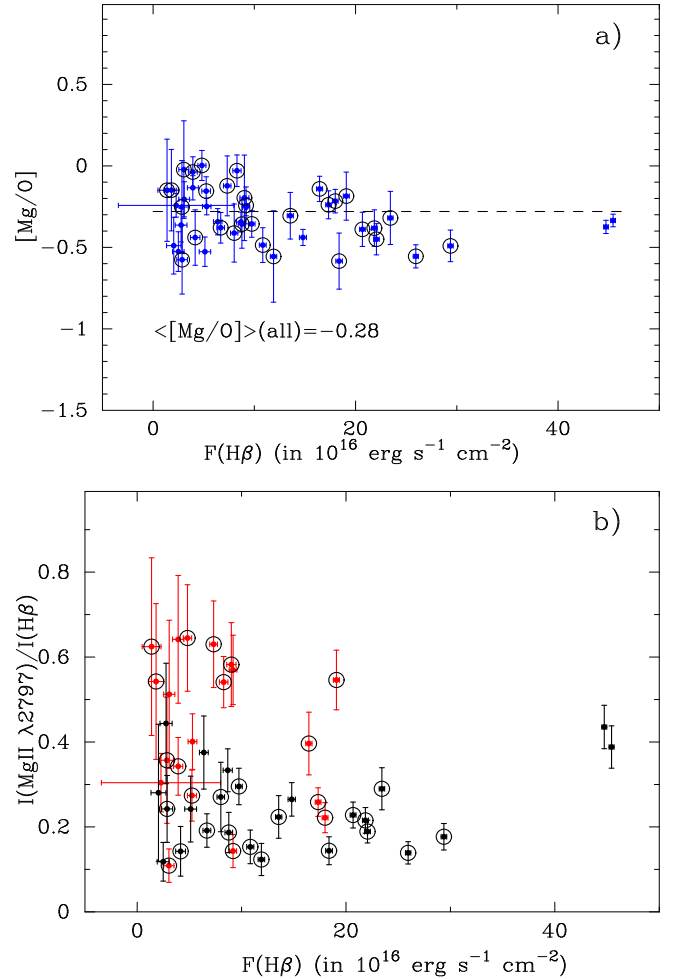

\hspace*{0.2cm}\psfig{figure=aa21010-12_fig7a.ps,angle=-90,width=8.5cm}
\vspace{0.3cm}
\hspace*{0.2cm}\psfig{figure=aa21010-12_fig7b.ps,angle=-90,width=8.5cm}
\caption{Dependence of the magnesium-to-oxygen abundance ratio relative
to the solar value [Mg/O] (a) and 
the $I$(Mg~{\sc ii} $\lambda$2797)/$I$(H$\beta$) ratio (b)
on observed fluxes of the H$\beta$ 
emission line corrected for both the Milky Way and internal extinctions. 
(a) All sample galaxies are shown.
(b)  The same as in (a) but H {\sc ii} regions with [Mg/O] $\geq$ --0.3 
being shown in red and those with [Mg/O] $<$ --0.3 in black. 
  The galaxies with measured  
[O~{\sc iii}]$\lambda$4363\AA\ line are encircled.
}
\label{fig7}
\end{figure}

\setcounter{table}{5}

\begin{table*}
\caption{Magnesium abundances collected from the literature \label{tab5}}
\begin{tabular}{cclrcl} \hline \hline
& N (item)  & [X/Y]$^a$                  & Value & reference & Comments  \\ 
(1) & (2) & (3) & (4) & (5) & (6) \\ \hline
& 1   & [Mg/H]$^c$ & $-$1.22 & 1 & Cold Galactic clouds  \\
& 2   & [Mg/H]$^c$ & $-$0.62 & 1 & Warm Galactic clouds  \\
& 3   & [Mg/O] & $-$1.00 & 1 & Cold Galactic clouds  \\
& 4   & [Mg/O] & $-$0.40 & 1 & Warm Galactic clouds  \\
& 5   & [Mg/H]$^c$ & $-$0.37 & 1 & Weaker low-velocity (WLV) absorption of interstellar clouds toward 23 Ori (\textit{HST}) \\
&6 & [Mg/H]$^c$ & $-$0.74 & 1 & Strong low-velocity (SLV) absorption in H {\sc ii} gas toward 23 Ori (\textit{HST}) \\
& 7   & [Mg/S] & $-$0.51 & 1 & WLV absorption of interstellar clouds toward 23 Ori (\textit{HST}) \\ 
& 8   & [Mg/S] & $-$0.89 & 1 & SLV absorption in H {\sc ii} gas toward 23 Ori (\textit{HST}) \\ 
& 9   & [Mg/H]$^c$ & $-$0.46 & 2 & Mean value of ISM toward 70 early type stars (\textit{Copernicus}) \\
& 10  & [Mg/H]$^c$ & $-$0.24 & 3 & Interstellar cloud \textit{A} in the direction of $\zeta$ Oph  \\   
LISM$^b$& 11  & [Mg/H] & $-$0.90 & 3 & Interstellar cloud \textit{B} in the direction of $\zeta$ Oph  \\ 
& 12  & [Mg/O] & $-$0.48 & 3 & ISM, \textit{A} component in the direction of $\zeta$ Oph  \\ 
& 13  & [Mg/O] & $-$0.69 & 3 & ISM, \textit{B} component in the direction of $\zeta$ Oph  \\
& 14  & [Mg/O] & $-$1.21 & 4 & $\alpha$ CMa A, 1st component of LISM within 100 pc (\textit{HST}) \\
& 15  & [Mg/O] & $-$0.61 & 4 & $\alpha$ CMa A, 2nd component of LISM within 100 pc (\textit{HST}) \\
& 16  & [Mg/O] & $-$0.89 & 4 & G191-B2B 1st component of LISM within 100 pc (\textit{HST}) \\
& 17  & [Mg/O] & $+$0.28 & 4 & G191-B2B 2nd component of LISM within 100 pc (\textit{HST}) \\
& 18  & [Mg/O] & $-$0.46 & 4 & Weighted mean of LISM within 100 pc (\textit{HST}) \\
& 19  & [Mg/H]$^c$ & $-$0.9  & 5 & $\zeta$ Pyx (observations of ISM toward 8 cool (G4-K0) giants, \textit{IUE}) \\ 
& 20  & [Mg/H]$^c$ & $-$0.9  & 5 & HD 81101 (\textit{IUE}) \\
& 21  & [Mg/H]$^c$ & $-$0.8  & 5 & $\zeta$ Vol, 1st component (\textit{IUE}) \\
& 22  & [Mg/H]$^c$ & $-$0.9  & 5 & $\zeta$ Vol, 2nd component (\textit{IUE}) \\
\\ \hline
& 23  & [Mg/H]$^c$ & $+$0.023& 6 & averaged values for 25 planetary nabulae (PNe), Mg {\sc ii} $\lambda$4481 \\
& 24  & [Mg/H]$^c$ & 0.0   & 7 & planetary nebula IC 418, (\textit{IUE}), Mg {\sc ii} $\lambda$$\lambda$2797,2803 \\
PNe$^d$ & 25  & [Mg/H]$^c$ & $-$1.1  & 7 & planetary nebula NGC 2440, (\textit{IUE}), Mg {\sc ii} $\lambda$$\lambda$2797,2803 \\
& 26  & [Mg/H]$^c$ & $-$0.64 & 7 & planetary nebula IC 4997, (\textit{IUE}), Mg {\sc ii} $\lambda$$\lambda$2797,2803 \\
\\ \hline
& 27 & [Mg/O]$^e$ & $-$0.94  & 8 & Mg {\sc ii} $\lambda$4481 line, based on observation of Orion by \citet{Esteban2004} \\
H {\sc ii}& 28 & [Mg/O] & $-$1.00 & 8 & Mg {\sc ii} $\lambda$4481 line, based on observation of Orion by \citet{Esteban2004} \\
regions   & 29 & [Mg/O] & $-$0.35 & 8 & 30 Dor, based on Mg {\sc ii} $\lambda$4481 and Mg {\sc i} $\lambda$$\lambda$4561,4571 \\
& 30 & [Mg/O]$^f$ & $-$0.67 & 8 & 30 Dor, based on Mg {\sc ii} $\lambda$4481 and Mg {\sc i} $\lambda$$\lambda$4561,4571 \\
\\ \hline
& 31 & [Mg/S]$^g$ & $-$0.09 & 9 & gas-phase abundances obtained in absorption system along the GRB HG 050820 \\ 
& 32 & [Mg/S] & $-$0.18 & 9 & absorption system toward the GRB HG 050820 \\ 
& 33 & [Mg/S] & 0.0   & 10 & absorption system toward the Q 0100+13 (z$_{abs}$=2.309) \\ 
QSO-& 34 & [Mg/S] & $-$0.04 & 11 & absorption system toward the Q B0841+129 (z$_{DLA}$=2.375) \\ 
-DLA-& 35 & [Mg/S] & $+$0.03 & 11 & absorption system toward the Q B0841+129 (z$_{DLA}$=2.476) \\ 
-GRB& 36 & [Mg/S] & $-$0.07 & 12 & absorption system toward the Q 1101--264 (z$_{abs}$=1.838) \\  
& 37 & [Mg/S] & $-$0.04 & 13 & mean value for absorption system toward the Q 1331+17 (z$_{abs}$=1.776) \\
& 38  & [Mg/S] & $+$0.13 & 14 & data for $\sim$85 DLA systems (Keck spectra) FJ 0812+32 (z$_{abs}$=2.6263) \\
& 39  & [Mg/S] & $-$0.49 & 14 & data for $\sim$85 DLA systems (Keck spectra) J 0900+42 (z$_{abs}$=3.2458) \\
\\ \hline
\end{tabular}

References. [1] \citet{Welty1999}, [2] \citet{Murray1983}, [3] \citet{Savage1992}, [4] \citet{Redfield2004}, 
[5] \citet{Molaro1986}, [6] \citet{Dinerstein2012}, [7] \citet{M88}, [8] \citet{PP10}, [9] \citet{Prochaska2007a}, 
[10] \citet{DZavadsky2004}, [11] \citet{DZavadsky2007}, [12] \citet{DZavadsky2003}, [13] \citet{DZavadsky2006}, 
[14] \citet{Prochaska2007}

$^a$All data were recalculated with solar abundances by \citet{A09}.

$^b$Local Interstellar Medium.

$^c$The solar abundance pattern is adopted for the local interstellar medium.

$^d$Planetary nebulae.

$^e$Adopted solar vicinity ISM abundances, average from B stars and solar 
neighbourhood abundances corrected 
for Galactic chemical evolution since the Sun was formed \citep{PP10}. 

$^f$Adopted abundances by \citet{Tsamis2005}. 

$^g$Solar abundances by \citet{Asplund2005}  are assumed, following \citet{Prochaska2007a}.
\end{table*}

\section{Results and discussion \label{S4}}

  In Figs. \ref{fig5}a and \ref{fig5}b the dependences 
of $[$Mg/O$]$ = log(Mg/O) -- log(Mg/O)$_\odot$ on 12+logO/H are shown
for the 45 spectra where 
the Mg~{\sc ii} emission lines was detected. 
  We note also that in many cases blue-shifted
Mg~{\sc ii} absorption near the Mg~{\sc ii} $\lambda$2797\AA\ emission line 
is present which may indicate of stellar
winds from cool massive stars, such as LBVs (see, for instance, 
J0207+0047,
J0925+2709, J1045+3225 in Fig. \ref{fig2}). 
  The encircled galaxies are those with no clear presence of blue-shifted 
absorption profiles (Fig. \ref{fig5}a).
  There is no obvious offset of the encircled points as compared 
to other data.
  This testifies to the correctness  of Mg {\sc ii} $\lambda$2797 and 
$\lambda$2803 line measurements in cases when the continuum is placed 
at the bottom
of absorption profiles.
  In Fig. \ref{fig5}b we divide the sample between objects with bright
Mg {\sc ii} lines ($I$($\lambda$2797\AA)/$I$(H$\beta$) $\geq$ 0.3 
and weak Mg {\sc ii} lines 
($I$($\lambda$2797\AA)/$I$(H$\beta$) $<$ 0.3. 
  Encircled galaxies in (b) are those where 
[O~{\sc iii}]$\lambda$4363\AA\ emission is detected.
 A slight shift of H {\sc ii} regions with bright Mg {\sc ii} emission 
line $\lambda$2797
to lower metallicity is seen in Fig. \ref{fig5}b. 

  Table \ref{tab6} summarises spectroscopic parameters for all 65 spectra, where
we show the galaxy name, the oxygen abundance 12+logO/H, the quantity
$[$Mg/O$]$, and the equivalent width
EW(H$\beta$) of the H$\beta$ emission line in \AA. We adopt 
log(Mg/O)$_\odot$ = --1.09 \citep{A09}. 
  We note that both O and Mg are subject to 
depletion onto dust. However, oxygen is by a factor of $\sim$ 10
more abundant than magnesium. 
 Therefore, the quantity
$[$Mg/O$]$ is mainly a characteristic of the magnesium depletion.

  The dependences of $[$Mg/O$]$ on EW(H$\beta$)
are shown in Fig. \ref{fig6}.
 Three H {\sc ii} regions from our sample have EW(H$\beta$) $>$ 250\AA\ and 
are outside of the figure. 
  Their $[$Mg/O$]$ values 
range between --0.02 and --0.24 (see Table \ref{tab6}), similar to  
the range of $[$Mg/O$]$ for the entire sample. 
  The galaxies with measured  
[O~{\sc iii}]$\lambda$4363\AA\ emission line are encircled.
  There is no trend in $[$Mg/O$]$ with EW(H$\beta$)
in Fig. \ref{fig6}. The EW(H$\beta$) is a characteristic
of the starburst age, hence of the excitation parameter $x$.
  Therefore, the absence of any correlation in Fig. \ref{fig6}  
indicates the correctness of the correction for unseen stages of Mg ionisation.

 To additionally check the accuracy of our measurements and our magnesium
and oxygen abundance determinations, we show in Fig. \ref{fig7}
the dependence of the magnesium-to-oxygen abundance ratio relative
to solar value $[$Mg/O$]$ (Fig. \ref{fig7}a) and 
$I$(Mg~{\sc ii} $\lambda$2797)/$I$(H$\beta$) (Fig. \ref{fig7}b)
on the observed fluxes of H$\beta$ 
emission lines corrected for both the Milky Way and internal extinctions 
for all sample galaxies. 
   The galaxies with measured  
[O~{\sc iii}]$\lambda$4363\AA\ emission line are encircled.
No differences in the distributions of the galaxies with measured
[O~{\sc iii}]$\lambda$4363\AA\ line and without this line 
is found in Figs. \ref{fig5}b, \ref{fig6}, and  \ref{fig7}a,b.
Thus, in the following discussion we use all the data together obtained with
the direct and semi-empirical methods.
  There is also no clear trend in $[$Mg/O$]$  and  
$I$(Mg~{\sc ii} $\lambda$2797)/$I$(H$\beta$) with corrected fluxes of H$\beta$.
 This is also evidence for the accuracy of our determinations.
 
  We do not find any obvious trend in $[$Mg/O$]$ with
oxygen abundance in the entire range of 12+logO/H = 7.52 -- 8.44
(Fig. \ref{fig5}).
  The mean value of the Mg/O ratio is by a factor of $\sim$2 lower than
the solar value, implying moderate Mg depletion of $\sim$50\%
in the dust phase. 
This depletion is significantly higher than the mean value 
obtained by \citet{Dinerstein2012} for PNe. 
 They derived Mg/O  
close to the solar value using the
recombination Mg~{\sc ii} $\lambda$4481 emission line. 
 These differences 
can be explained by the known 
discrepancy between values derived from recombination lines (RL)
and collisionally excited lines (CEL)
\citep{Esteban2009,Peimbert2005,Davey2000,Guseva2011}.
  Abundances obtained from RLs tend to be higher than those derived from 
CEL lines.
   The origin of the nebular abundance discrepancy problem 
is currently not known. 
  On the other hand, \citet{Barlow2003}
and \citet{Wang2007} do not empirically find any such discrepancy for Mg/H.
 Moreover, the Mg fraction in dust also obtained using the
recombination Mg~{\sc ii} $\lambda$4481 emission line is much higher in 
30 Doradus (72\%) and in the Orion nebula (91\%) \citep{PP10}
than in the PNe analysed by \citet{Dinerstein2012}.

  The negative values of $[$Mg/O$]$ in our galaxies can be due to the 
absorption of the Mg~{\sc ii} $\lambda$2797, $\lambda$2803 emission by the 
interstellar gas outside the H~{\sc ii} regions.
The spectral resolution of the SDSS spectra is insufficient to separate 
the Mg~{\sc ii} absorption and emission to estimate the effect of 
the interstellar absorption and to correct emission-line intensities
for this effect. 
  An additional source of uncertainties may arise due to the presence
of broad blue-shifted absorption lines in many spectra of our sample
superposed with the emission lines
(Fig. \ref{fig2}). 
  One of the most evident cases is the galaxy J1045$+$3225.
These lines are likely to be broad lines with P Cygni profiles produced
by cool massive stars (red supergiants, LBVs) with a stellar wind
\citep[e.g. ][]{H01}. 
  The presence of these lines will introduce uncertainties into the 
placement of the continuum for measurements of the intensities of 
nebular emission lines.
    We measure magnesium emission lines placing
the continuum at the bottom of the absorption lines 
when it is obviously seen
in the observed spectra as described above.
In spectra of some other galaxies 
there is no clear evidence of the 
blue-shifted broad absorption (see Fig. \ref{fig2}). 
  To verify the effect of this absorption
on the derived Mg abundance, we 
divided the objects in Fig. \ref{fig5}a into two samples, those with and those
without evident blue-shifted broad absorption. It is very important that 
the encircled galaxies without blue-shifted absorption 
have the same $[$Mg/O$]$ and dispersion as other galaxies. 

\begin{figure*}
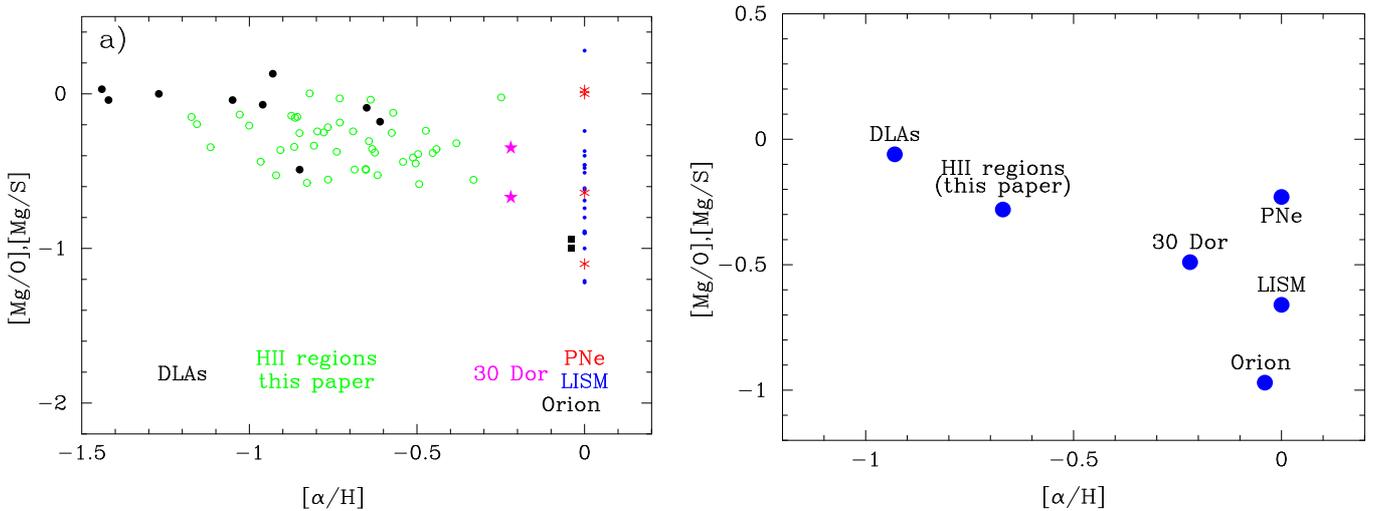

\hspace*{0.0cm}\psfig{figure=aa21010-12_fig8a.ps,angle=-90,width=8.5cm,clip=}
\hspace*{0.4cm}\psfig{figure=aa21010-12_fig8b.ps,angle=-90,width=8.9cm,clip=}
\caption{(a) Comparison of magnesium depletion for data available  
from the literature (denoted by different colours and symbols)
and (b) comparison of the magnesium depletion averaged for objects of each of four types
(LISM (Local interstellar medium), H {\sc ii} regions, our Mg {\sc ii} sample 
(this paper), and DLA-QSO-GRB 
(Damped Ly$\alpha$ absorption -- Quasi-stellar object -- Gamma-ray burst) 
absorption systems.
[$\alpha$/H] is [O/H] for 
LISM, H {\sc ii} regions, and our sample galaxies, and $[$S/H$]$ 
for DLA-QSA-GRB absorption systems.
}
\label{fig8}
\end{figure*}

Summarising, we find that the Mg depletion in our galaxies is probably present
but at a relatively low level with roughly half of the Mg in dust.
 This level of Mg depletion onto dust can be considered as an upper limit
because the interstellar absorption is not known and is
difficult to be taken it into account.
  For comparison, \citet{Jenkins2009} found
the relative proportions of different elements that are incorporated into dust
at different stages of grain growth based on gas-phase element 
abundances for 17 different elements over
243 sight lines in the local part of our Galaxy. 
According to them $\sim$70\% 
\citep[$\sim$50\% on Fig. 5 by ][]{Jenkins2009} of magnesium is 
locked onto dust grains and $\sim$95\%
\citep[$\sim$90\% on Fig. 7 by ][]{Jenkins2009} of iron is also in 
solid form \citep[see also Fig. 1 by ][]{Jenkins2009}.

\subsection{Comparison of magnesium depletion in objects of different types \label{comp}}

 We compared magnesium depletion obtained for our low-metallicity, emission-line 
star-forming galaxies with available data from the literature for
objects of different types.
  There are many DLA and QSO absorption systems with magnesium abundance 
determination from Mg {\sc ii} $\lambda$2797,$\lambda$2803 lines.  
 However, only an upper limit to the 
Mg{\sc ii} abundance 
has been obtained in the majority of the cases. 
  We include in Table \ref{tab5} only those objects for which the exact values 
of number densities for hydrogen, magnesium, and oxygen (or sulphur) have been 
collected. We recalculated all data with solar abundances
by \citet{A09} where it was necessary.
  To be more specific, we display in Fig. \ref{fig8}a all data from 
Table \ref{tab5}, indicating the different type of objects by different 
colours and symbols. 
   The  value of [$\alpha$/H] is in terms of [S/H] for 
DLA-QSA-GRB absorption systems and
[O/H] for LISM, H {\sc ii} regions, and our sample galaxies.

 Following \citet{Lebouteiller2008} we adopted the oxygen abundance of 30 Dor
in the LMC to be a factor of $\sim$0.6 lower than the solar value. 
  The mean oxygen abundance of our 
Mg sample
is 12 + logO/H = 8.02 for 45 H {\sc ii} regions. 
  A solar abundance of 
12 + logO/H = 8.69 \citep{A09} was adopted for the local interstellar
medium (clouds toward Galactic stars and PNe). 
 Following \citet{PP10} an oxygen abundance of 12 + logO/H = 8.65
was adopted for the Orion nebula \citep[determination 
based on the observations by][]{Esteban2004}.
  In all comparisons for DLA absorption systems we used 
the non-refractory element sulphur for 
metallicity estimation and for the determination of magnesium depletion
because oxygen absorption lines are generally  
saturated \citep[e.g.][]{Prochaska2007}.  

  In Fig. \ref{fig8}b we compare [Mg/O] and [$\alpha$/H] averaged 
for each of four types of objects (LISM, H {\sc ii} regions, our  
sample,  and DLA-QSO-GRB absorption systems).
  The averaged values of [Mg/O] and [$\alpha$/H] are obtained
from the logarithms of the average $<$Mg/O$>$ and $<$$\alpha$/H$>$.

  This is to be compared to iron depletion. 
\citet{Rodriguez2005} discovered 
the trend of increasing Fe depletion at higher oxygen abundance.
For the Galactic H {\sc ii} regions and PNe they obtained high Fe depletion, 
with fewer than 5\% of their Fe atoms in 
the gas phase, whereas the 
metal-deficient blue compact galaxy SBS 0335--052 could have only from 13\%
to 40\% of Fe in the gas phase. 
  \citet{I06} also obtained the trend in Fe depletion with metallicity.
Considering the iron depletion in low-metallicity, star-forming
emission-line galaxies they 
find that $\sim$80\% of iron is confined in dust
in galaxies with 12+logO/H $\sim$ 8.0 (their Fig. 11l).

   Despite  a wide spread of points for the individual objects 
(Fig. \ref{fig8}a), there is also a clear trend in averaged $[$Mg/O$]$ 
with metallicity
in Fig. \ref{fig8}b for different types of objects, indicating that
the depletion is higher for higher metallicity objects.
  Thus, we confirm the previous findings by \citet{Jenkins2009}
 of increasing Mg depletion with increasing metallicity. 

\section{Conclusions \label{S5}}

We have presented 65 SDSS spectra of low-metallicity, emission-line star-forming
galaxies with redshifts $z$ $\sim$ 0.36 -- 0.70, with the aim of studing the
interstellar magnesium abundance as derived from the resonance 
emission-line doublet Mg~{\sc ii} $\lambda$2797, $\lambda$2803. This 
emission is detected in 45 galaxies, or in more than two thirds of the sample.
Our main results are as follows:

1. Using the grid of 576 CLOUDY photoionised H~{\sc ii} region models, we 
obtained fits of the electron temperature $T_e$(Mg~{\sc ii}) in the Mg$^+$
zone as a function of the electron temperature $T_e$(O~{\sc ii}) in the
O$^+$ zone. Furthermore, using the same CLOUDY models we obtained fits
of the ionisation correction factor $ICF$(Mg$^+$) as a function of
the excitation parameter $x$ = O$^{2+}$/O.

2. We derived oxygen and magnesium abundances. 
The [O~{\sc iii}] $\lambda$4363 emission line is seen  
in $\sim$63\% of the sources of the total sample and $\sim$70\% 
of the galaxies in which the Mg~{\sc ii} lines were detected,
allowing an accurate abundance determination
with the direct $T_e$ method. The element abundances in the
remaining galaxies we are derived using a semi-empirical method that is
based on strong nebular oxygen lines. 
  Data obtained with the two methods are indistinguishable in all 
distributions.
The oxygen abundances 
12+logO/H for the sample galaxies are in the relatively wide range from
7.52 to 8.44.

3. We found that the Mg/O abundance ratio in our galaxies is lower by a factor
of two than in the Sun, implying a moderate Mg depletion onto
dust with $\sim$50\% of magnesium confined in dust. This level of depletion
is higher than in high-redshift, low-metallicity damped Ly$\alpha$
(DLA) systems with measured Mg abundances. 
  However, it is much lower than 
in H~{\sc ii} regions with higher metallicity (e.g., in the Orion nebula and
30 Doradus, where the fraction of Mg in dust is 91\% and 72\%, respectively),
in PNe (averaged value) and in interstellar clouds
of the LISM.

4. Despite a wide spread of points for the individual determinations 
there is a clear trend in $[$Mg/O$]$ or  $[$Mg/S$]$ with metallicity
when averaged magnesium abundance and metallicity for different types of 
objects is used.
This implies that the magnesium depletion is higher in higher metallicity 
objects.

\acknowledgements

N.G.G. and Y.I.I. acknowledge the hospitality of the Max-Planck 
Institute for Radioastronomy, Bonn, Germany.   
  This research made use of the NASA/IPAC Extragalactic Database
(NED), which is operated by the Jet Propulsion Laboratory, California
Institute of Technology, under contract with the National Aeronautics 
and Space Administration.
    Funding for the Sloan Digital Sky Survey (SDSS), and SDSS-II has been 
provided by the Alfred P. Sloan Foundation, the Participating Institutions, 
the National Science Foundation, the U.S. Department of Energy, the National 
Aeronautics and Space Administration, the Japanese Monbukagakusho, and the 
Max Planck Society, and the Higher Education Funding Council for England. 
 

\Online

\renewcommand{\baselinestretch}{1.0}

\newpage

\setcounter{table}{1}

 \begin{longtable}{lccccccc}
 \caption{Extinction-corrected emission-line intensities}
 \label{tab2} \\ \hline \hline
 \endfirsthead
 \caption{---{\sl Continued.}} \\
 \hline\hline
Ion&$I$($\lambda$)/$I$(H$\beta$)&$I$($\lambda$)/$I$(H$\beta$)&$I$($\lambda$)/$I$(H$\beta$)&$I$($\lambda$)/$I$(H$\beta$)&$I$($\lambda$)/$I$(H$\beta$)&$I$($\lambda$)/$I$(H$\beta$)&$I$($\lambda$)/$I$(H$\beta$)\\
 \cline{2-8}
  &\multicolumn{7}{c}{\sc Galaxy} \\  \hline
 \endhead
 \hline
 \endfoot
Ion&$I$($\lambda$)/$I$(H$\beta$)&$I$($\lambda$)/$I$(H$\beta$)&$I$($\lambda$)/$I$(H$\beta$)&$I$($\lambda$)/$I$(H$\beta$)&$I$($\lambda$)/$I$(H$\beta$)&$I$($\lambda$)/$I$(H$\beta$)&$I$($\lambda$)/$I$(H$\beta$)\\
 \cline{2-8}
  &\multicolumn{7}{c}{\sc Galaxy} \\  \hline
 &J0006$-$0903 &J0101$-$0057 &J0104$+$0013 &J0142$+$0108 &J0149$+$0100 &J0155$-$0107 &J0207$+$0047 \\ \hline
2797 Mg {\sc ii}        &0.55$\pm$0.07&0.54$\pm$0.06&0.63$\pm$0.10&0.22$\pm$0.05&~~...&0.15$\pm$0.04&0.26$\pm$0.03\\
2804 Mg {\sc ii}        &0.39$\pm$0.08&0.43$\pm$0.05&0.46$\pm$0.10&0.18$\pm$0.05&~~...&0.07$\pm$0.03&0.16$\pm$0.03\\
3727 [O {\sc ii}]       &3.00$\pm$0.10&2.08$\pm$0.10&3.09$\pm$0.14&1.80$\pm$0.07&1.73$\pm$0.08&1.39$\pm$0.07&1.66$\pm$0.06\\
3868 [Ne {\sc iii}]     &0.33$\pm$0.03&0.41$\pm$0.04&0.50$\pm$0.06&0.31$\pm$0.03&0.34$\pm$0.03&0.50$\pm$0.04&0.37$\pm$0.02\\
4101 H$\delta$          &0.30$\pm$0.03&0.25$\pm$0.03&0.32$\pm$0.05&0.26$\pm$0.03&0.25$\pm$0.03&0.26$\pm$0.03&0.25$\pm$0.02\\
4340 H$\gamma$          &0.47$\pm$0.03&0.48$\pm$0.04&0.46$\pm$0.04&0.47$\pm$0.03&0.48$\pm$0.03&0.47$\pm$0.03&0.47$\pm$0.02\\
4363 [O {\sc iii}]      &0.04$\pm$0.02&0.09$\pm$0.02&0.05$\pm$0.02&0.04$\pm$0.01&0.07$\pm$0.02&0.10$\pm$0.01&0.05$\pm$0.01\\
4861 H$\beta$           &1.00$\pm$0.03&1.00$\pm$0.04&1.00$\pm$0.04&1.00$\pm$0.03&1.00$\pm$0.04&1.00$\pm$0.03&1.00$\pm$0.03\\
4959 [O {\sc iii}]      &1.02$\pm$0.03&1.68$\pm$0.05&1.39$\pm$0.04&1.24$\pm$0.03&1.49$\pm$0.05&2.11$\pm$0.06&1.85$\pm$0.04\\
5007 [O {\sc iii}]      &2.96$\pm$0.07&5.08$\pm$0.12&4.20$\pm$0.10&3.69$\pm$0.08&4.69$\pm$0.11&6.30$\pm$0.14&5.34$\pm$0.11\\
$C$(H$\beta$)(MW)       &  0.05&  0.04&  0.04&  0.04&  0.03&  0.04&  0.03\\
$C$(H$\beta$)(int)      &  0.50&  0.00&  0.50&  0.22&  0.00&  0.01&  0.00\\
EW$_{\rm abs}$          &  1.11&  0.26&  0.88&  1.13&  1.11&  2.13&  1.36\\
EW(H$\beta$)            & 50.63& 64.48& 49.76& 68.95& 94.39&185.30&107.30\\
$F$(H$\beta$)$^a$       & 19.08&  8.29&  7.32& 13.55&  7.26& 10.83& 17.33\\ \hline
 &J0232$-$0021 &J0301$-$0806 &J0319$-$0104 &J0340$-$0048 &J0747$+$3147 &J0749$+$3142 &J0756$+$3057 \\ \hline
2797 Mg {\sc ii}        &0.14$\pm$0.03&0.27$\pm$0.06&0.30$\pm$0.04&0.30$\pm$0.07&~~...&0.64$\pm$0.13&0.64$\pm$0.15\\
2804 Mg {\sc ii}        &0.09$\pm$0.03&0.24$\pm$0.06&0.16$\pm$0.04&0.28$\pm$0.08&~~...&0.42$\pm$0.11&0.49$\pm$0.14\\
3727 [O {\sc ii}]       &2.02$\pm$0.07&1.31$\pm$0.08&2.18$\pm$0.08&2.02$\pm$0.15&3.46$\pm$0.31&2.02$\pm$0.10&2.97$\pm$0.19\\
3868 [Ne {\sc iii}]     &0.31$\pm$0.03&0.61$\pm$0.05&0.45$\pm$0.03&0.45$\pm$0.06&0.37$\pm$0.10&0.49$\pm$0.04&~~~~...\\
4101 H$\delta$          &0.25$\pm$0.03&0.34$\pm$0.04&0.26$\pm$0.02&0.42$\pm$0.21&~~...&0.27$\pm$0.03&0.32$\pm$0.08\\
4340 H$\gamma$          &0.47$\pm$0.03&0.53$\pm$0.04&0.47$\pm$0.02&0.77$\pm$2.21&0.40$\pm$0.08&0.48$\pm$0.03&0.45$\pm$0.06\\
4363 [O {\sc iii}]      &0.03$\pm$0.01&0.15$\pm$0.02&0.06$\pm$0.01&~~...&~~...&0.09$\pm$0.02&~~~~...\\
4861 H$\beta$           &1.00$\pm$0.03&1.00$\pm$0.04&1.00$\pm$0.03&1.00$\pm$2.57&1.00$\pm$0.08&1.00$\pm$0.04&1.00$\pm$0.05\\
4959 [O {\sc iii}]      &1.30$\pm$0.04&2.14$\pm$0.07&1.57$\pm$0.04&1.43$\pm$0.07&0.88$\pm$0.07&1.65$\pm$0.05&0.64$\pm$0.04\\
5007 [O {\sc iii}]      &3.92$\pm$0.09&6.37$\pm$0.16&4.84$\pm$0.11&4.13$\pm$0.16&2.51$\pm$0.14&4.69$\pm$0.12&1.81$\pm$0.07\\
$C$(H$\beta$)(MW)       &  0.03&  0.09&  0.08&  0.12&  0.08&  0.08&  0.07\\
$C$(H$\beta$)(int)      &  0.00&  0.00&  0.07&  0.00&  0.50&  0.41&  0.50\\
EW$_{\rm abs}$          &  3.51&  0.00&  1.88&  0.00&  4.99&  0.79&  1.73\\
EW(H$\beta$)            & 82.57&167.60& 94.19&199.30& 78.63&147.50& 50.28\\
$F$(H$\beta$)$^a$       & 18.37&  5.25&  9.74&  2.28&  2.08&  4.82&  3.93\\ \hline
 &J0757$+$3148 &J0804$+$1748 &J0807$+$4002 &J0812$+$3200 &J0812$+$3200 &J0813$+$3339 &J0815$+$3129 \\ \hline
2797 Mg {\sc ii}        &0.34$\pm$0.07&0.19$\pm$0.04&~~...&~~...&0.14$\pm$0.06&0.51$\pm$0.17&~~~~...\\
2804 Mg {\sc ii}        &0.23$\pm$0.06&0.15$\pm$0.04&~~...&~~...&0.10$\pm$0.06&0.43$\pm$0.18&~~~~...\\
3727 [O {\sc ii}]       &1.30$\pm$0.08&1.75$\pm$0.08&2.45$\pm$0.08&1.53$\pm$0.09&1.47$\pm$0.10&2.86$\pm$0.19&2.08$\pm$0.16\\
3868 [Ne {\sc iii}]     &0.47$\pm$0.05&0.48$\pm$0.04&0.27$\pm$0.02&0.42$\pm$0.05&0.48$\pm$0.05&0.23$\pm$0.07&0.17$\pm$0.07\\
4101 H$\delta$          &0.25$\pm$0.03&0.26$\pm$0.03&0.30$\pm$0.03&0.34$\pm$0.20&0.27$\pm$0.04&0.27$\pm$0.06&0.26$\pm$0.07\\
4340 H$\gamma$          &0.48$\pm$0.03&0.47$\pm$0.03&0.47$\pm$0.02&0.51$\pm$0.04&0.47$\pm$0.04&0.48$\pm$0.06&0.47$\pm$0.07\\
4363 [O {\sc iii}]      &0.09$\pm$0.02&0.07$\pm$0.01&0.06$\pm$0.01&0.10$\pm$0.02&0.07$\pm$0.02&~~...&~~~~...\\
4861 H$\beta$           &1.00$\pm$0.04&1.00$\pm$0.04&1.00$\pm$0.03&1.00$\pm$0.04&1.00$\pm$0.05&1.00$\pm$0.05&1.00$\pm$0.06\\
4959 [O {\sc iii}]      &2.10$\pm$0.07&1.78$\pm$0.05&1.22$\pm$0.03&1.94$\pm$0.06&2.06$\pm$0.07&0.66$\pm$0.04&0.87$\pm$0.05\\
5007 [O {\sc iii}]      &6.16$\pm$0.17&5.31$\pm$0.13&3.53$\pm$0.08&5.77$\pm$0.15&6.09$\pm$0.16&2.15$\pm$0.08&2.54$\pm$0.10\\
$C$(H$\beta$)(MW)       &  0.06&  0.03&  0.07&  0.05&  0.05&  0.05&  0.05\\
$C$(H$\beta$)(int)      &  0.00&  0.09&  0.50&  0.00&  0.02&  0.49&  0.23\\
EW$_{\rm abs}$          &  1.32&  0.75&  2.49&  0.00&  0.00&  0.00&  0.51\\
EW(H$\beta$)            &147.20&165.20&120.90&339.70&116.20& 43.42& 25.24\\
$F$(H$\beta$)$^a$       &  3.93&  6.68& 23.45&  5.09&  4.16&  3.07&  3.36\\ \hline
 &J0817$+$3159 &J0837$+$3108 &J0838$+$3057 &J0844$+$3241 &J0855$+$0318 &J0904$+$2806 &J0914$+$4451 \\ \hline
2797 Mg {\sc ii}        &0.19$\pm$0.05&0.24$\pm$0.08&0.12$\pm$0.05&0.54$\pm$0.18&0.22$\pm$0.04&0.58$\pm$0.10&0.14$\pm$0.03\\
2804 Mg {\sc ii}        &0.16$\pm$0.05&0.10$\pm$0.08&0.12$\pm$0.06&0.54$\pm$0.19&0.07$\pm$0.03&0.26$\pm$0.09&0.08$\pm$0.02\\
3727 [O {\sc ii}]       &1.82$\pm$0.09&2.46$\pm$0.16&1.72$\pm$0.13&2.95$\pm$0.24&0.88$\pm$0.04&3.08$\pm$0.15&1.51$\pm$0.05\\
3868 [Ne {\sc iii}]     &0.47$\pm$0.04&0.33$\pm$0.06&0.49$\pm$0.07&0.38$\pm$0.08&0.49$\pm$0.03&0.19$\pm$0.05&~~~~...\\
4101 H$\delta$          &0.27$\pm$0.03&0.27$\pm$0.05&0.26$\pm$0.05&0.27$\pm$0.07&~~...&0.33$\pm$0.07&0.28$\pm$0.02\\
4340 H$\gamma$          &0.47$\pm$0.03&0.47$\pm$0.05&0.47$\pm$0.05&0.48$\pm$0.06&0.49$\pm$0.02&0.47$\pm$0.06&0.49$\pm$0.02\\
4363 [O {\sc iii}]      &0.05$\pm$0.02&0.07$\pm$0.03&~~...&0.06$\pm$0.04&0.11$\pm$0.01&0.07$\pm$0.03&0.12$\pm$0.01\\
4861 H$\beta$           &1.00$\pm$0.04&1.00$\pm$0.05&1.00$\pm$0.06&1.00$\pm$0.07&1.00$\pm$0.03&1.00$\pm$0.04&1.00$\pm$0.03\\
4959 [O {\sc iii}]      &1.89$\pm$0.05&1.14$\pm$0.06&1.89$\pm$0.08&0.92$\pm$0.07&1.98$\pm$0.05&0.47$\pm$0.03&1.97$\pm$0.04\\
5007 [O {\sc iii}]      &5.60$\pm$0.13&3.85$\pm$0.13&5.68$\pm$0.20&3.08$\pm$0.16&6.17$\pm$0.13&1.43$\pm$0.05&6.12$\pm$0.13\\
$C$(H$\beta$)(MW)       &  0.05&  0.06&  0.06&  0.04&  0.05&  0.04&  0.02\\
$C$(H$\beta$)(int)      &  0.17&  0.03&  0.12&  0.41&  0.00&  0.50&  0.00\\
EW$_{\rm abs}$          &  0.00&  0.00&  3.04&  3.03&  0.00&  2.33&  0.00\\
EW(H$\beta$)            &117.00& 63.50&113.90& 62.40&388.90& 31.03&170.70\\
$F$(H$\beta$)$^a$       &  8.77&  2.87&  2.49&  1.81& 17.99&  9.02& 25.95\\ \hline
 &J0915$+$3344 &J0918$+$3301 &J0918$+$3331 &J0922$+$3330 &J0925$+$2709 &J0956$+$3203 &J0957$+$3314 \\ \hline
2797 Mg {\sc ii}        &0.36$\pm$0.15&0.11$\pm$0.04&0.62$\pm$0.21&0.28$\pm$0.16&0.12$\pm$0.04&~~...&0.44$\pm$0.14\\
2804 Mg {\sc ii}        &0.38$\pm$0.16&0.07$\pm$0.04&0.39$\pm$0.19&0.22$\pm$0.16&0.10$\pm$0.04&~~...&0.32$\pm$0.14\\
3727 [O {\sc ii}]       &2.84$\pm$0.22&0.37$\pm$0.05&2.80$\pm$0.29&3.16$\pm$0.28&1.92$\pm$0.08&2.17$\pm$0.13&3.52$\pm$0.23\\
3868 [Ne {\sc iii}]     &0.31$\pm$0.08&0.27$\pm$0.04&0.16$\pm$0.09&0.34$\pm$0.10&0.31$\pm$0.03&0.20$\pm$0.05&0.24$\pm$0.08\\
4101 H$\delta$          &0.22$\pm$0.07&0.26$\pm$0.04&0.37$\pm$0.11&0.24$\pm$0.09&0.31$\pm$0.03&0.25$\pm$0.05&0.31$\pm$0.07\\
4340 H$\gamma$          &0.33$\pm$0.06&0.47$\pm$0.04&0.45$\pm$0.08&0.32$\pm$0.05&0.45$\pm$0.03&0.48$\pm$0.05&0.46$\pm$0.06\\
4363 [O {\sc iii}]      &0.04$\pm$0.03&0.03$\pm$0.01&0.09$\pm$0.05&~~...&0.02$\pm$0.01&0.06$\pm$0.03&~~~~...\\
4861 H$\beta$           &1.00$\pm$0.06&1.00$\pm$0.05&1.00$\pm$0.09&1.00$\pm$0.07&1.00$\pm$0.03&1.00$\pm$0.05&1.00$\pm$0.06\\
4959 [O {\sc iii}]      &1.13$\pm$0.06&1.81$\pm$0.07&0.71$\pm$0.07&1.31$\pm$0.08&1.31$\pm$0.04&0.82$\pm$0.04&0.69$\pm$0.05\\
5007 [O {\sc iii}]      &3.91$\pm$0.15&5.50$\pm$0.17&2.06$\pm$0.15&3.96$\pm$0.18&3.98$\pm$0.09&2.52$\pm$0.08&2.13$\pm$0.09\\
$C$(H$\beta$)(MW)       &  0.02&  0.02&  0.02&  0.02&  0.03&  0.02&  0.02\\
$C$(H$\beta$)(int)      &  0.50&  0.02&  0.50&  0.50&  0.04&  0.00&  0.50\\
EW$_{\rm abs}$          &  4.99&  0.00&  2.99&  0.23&  0.00&  1.89&  0.08\\
EW(H$\beta$)            &207.80&424.50& 61.38&131.50& 73.58& 43.63& 39.13\\
$F$(H$\beta$)$^a$       &  2.85&  3.04&  1.38&  2.04& 11.90&  5.00&  2.77\\ \hline
 &J1000$+$3324 &J1002$+$3228 &J1007$+$0446 &J1045$+$3225 &J1045$+$3225 &J1048$+$6359 &J1056$+$4758 \\ \hline
2797 Mg {\sc ii}        &~~...&~~...&~~...&0.19$\pm$0.03&0.22$\pm$0.03&0.33$\pm$0.05&0.14$\pm$0.04\\
2804 Mg {\sc ii}        &~~...&~~...&~~...&0.11$\pm$0.03&0.15$\pm$0.03&0.18$\pm$0.06&0.04$\pm$0.03\\
3727 [O {\sc ii}]       &2.71$\pm$0.23&2.31$\pm$0.18&1.83$\pm$0.06&1.90$\pm$0.06&2.01$\pm$0.07&2.00$\pm$0.10&0.56$\pm$0.04\\
3868 [Ne {\sc iii}]     &0.18$\pm$0.08&~~...&0.26$\pm$0.02&0.41$\pm$0.03&0.44$\pm$0.03&0.27$\pm$0.04&0.45$\pm$0.04\\
4101 H$\delta$          &0.26$\pm$0.07&0.21$\pm$0.07&0.38$\pm$0.02&0.26$\pm$0.02&0.27$\pm$0.02&0.26$\pm$0.04&0.25$\pm$0.03\\
4340 H$\gamma$          &0.47$\pm$0.07&0.49$\pm$0.07&0.44$\pm$0.02&0.49$\pm$0.02&0.47$\pm$0.02&0.47$\pm$0.04&0.55$\pm$0.03\\
4363 [O {\sc iii}]      &~~...&~~...&~~...&0.06$\pm$0.01&0.05$\pm$0.01&~~...&0.11$\pm$0.01\\
4861 H$\beta$           &1.00$\pm$0.08&1.00$\pm$0.06&1.00$\pm$0.03&1.00$\pm$0.03&1.00$\pm$0.03&1.00$\pm$0.04&1.00$\pm$0.03\\
4959 [O {\sc iii}]      &1.06$\pm$0.08&0.65$\pm$0.05&1.27$\pm$0.03&1.77$\pm$0.04&1.70$\pm$0.04&0.72$\pm$0.03&2.23$\pm$0.06\\
5007 [O {\sc iii}]      &3.05$\pm$0.16&1.96$\pm$0.10&3.70$\pm$0.08&5.33$\pm$0.11&5.18$\pm$0.11&2.24$\pm$0.07&6.81$\pm$0.15\\
$C$(H$\beta$)(MW)       &  0.02&  0.02&  0.02&  0.02&  0.02&  0.01&  0.02\\
$C$(H$\beta$)(int)      &  0.02&  0.00&  0.00&  0.00&  0.12&  0.00&  0.00\\
EW$_{\rm abs}$          &  0.07&  1.95&  0.00&  0.00&  0.00&  1.58&  0.00\\
EW(H$\beta$)            & 48.63& 50.22& 65.53& 77.87& 75.07& 41.96&431.90\\
$F$(H$\beta$)$^a$       &  2.15&  2.25& 23.75& 22.06& 21.86&  8.68&  9.18\\ \hline
 &J1108$+$6344 &J1112$+$6331 &J1128$-$0317 &J1135$+$6025 &J1136$-$0223 &J1213$+$2705 &J1408$+$0224 \\ \hline
2797 Mg {\sc ii}        &0.18$\pm$0.03&~~...&~~...&~~...&~~...&0.26$\pm$0.04&0.40$\pm$0.07\\
2804 Mg {\sc ii}        &0.08$\pm$0.03&~~...&~~...&~~...&~~...&0.20$\pm$0.04&0.31$\pm$0.07\\
3727 [O {\sc ii}]       &1.75$\pm$0.06&2.13$\pm$0.08&2.37$\pm$0.08&1.98$\pm$0.07&3.56$\pm$0.15&2.40$\pm$0.09&1.83$\pm$0.08\\
3868 [Ne {\sc iii}]     &0.32$\pm$0.02&0.37$\pm$0.03&0.33$\pm$0.03&0.20$\pm$0.02&0.49$\pm$0.06&0.18$\pm$0.03&0.52$\pm$0.04\\
4101 H$\delta$          &0.26$\pm$0.02&0.24$\pm$0.02&0.21$\pm$0.02&0.32$\pm$0.03&0.26$\pm$0.05&0.21$\pm$0.03&0.28$\pm$0.03\\
4340 H$\gamma$          &0.47$\pm$0.02&0.50$\pm$0.03&0.55$\pm$0.02&0.46$\pm$0.03&0.47$\pm$0.04&0.49$\pm$0.03&0.48$\pm$0.03\\
4363 [O {\sc iii}]      &0.05$\pm$0.01&0.12$\pm$0.01&0.06$\pm$0.01&0.03$\pm$0.01&~~...&~~...&0.10$\pm$0.02\\
4861 H$\beta$           &1.00$\pm$0.03&1.00$\pm$0.03&1.00$\pm$0.03&1.00$\pm$0.03&1.00$\pm$0.04&1.00$\pm$0.03&1.00$\pm$0.03\\
4959 [O {\sc iii}]      &1.42$\pm$0.03&1.62$\pm$0.04&1.39$\pm$0.04&0.61$\pm$0.02&1.09$\pm$0.04&0.94$\pm$0.03&1.58$\pm$0.04\\
5007 [O {\sc iii}]      &4.07$\pm$0.09&4.91$\pm$0.11&3.99$\pm$0.08&1.82$\pm$0.04&3.31$\pm$0.09&2.72$\pm$0.07&4.83$\pm$0.11\\
$C$(H$\beta$)(MW)       &  0.01&  0.01&  0.05&  0.02&  0.04&  0.03&  0.04\\
$C$(H$\beta$)(int)      &  0.09&  0.00&  0.00&  0.50&  0.27&  0.00&  0.44\\
EW$_{\rm abs}$          &  4.64&  0.05&  0.00&  2.76&  0.45&  0.74&  0.00\\
EW(H$\beta$)            &108.00&132.80& 32.54& 38.46& 38.27& 35.67&131.60\\
$F$(H$\beta$)$^a$       & 29.37& 14.33& 17.52& 43.44&  8.88& 14.80& 16.45\\ \hline
 &J1420$+$2628 &J1430$+$4802 &J1544$+$3308 &J1548$+$0727 &J1555$+$3543 &J1559$+$0634 &J1602$+$4002 \\ \hline
2797 Mg {\sc ii}        &~~...&~~...&~~...&0.23$\pm$0.03&0.29$\pm$0.05&0.38$\pm$0.09&~~~~...\\
2804 Mg {\sc ii}        &~~...&~~...&~~...&0.08$\pm$0.02&0.18$\pm$0.04&0.19$\pm$0.08&~~~~...\\
3727 [O {\sc ii}]       &1.99$\pm$0.06&3.28$\pm$0.15&2.09$\pm$0.11&1.74$\pm$0.06&2.27$\pm$0.08&2.38$\pm$0.11&2.11$\pm$0.10\\
3868 [Ne {\sc iii}]     &0.26$\pm$0.02&0.25$\pm$0.05&0.19$\pm$0.04&0.30$\pm$0.02&0.33$\pm$0.03&0.24$\pm$0.04&0.33$\pm$0.04\\
4101 H$\delta$          &0.26$\pm$0.02&0.26$\pm$0.07&0.26$\pm$0.05&0.27$\pm$0.02&0.29$\pm$0.02&0.22$\pm$0.06&0.34$\pm$0.03\\
4340 H$\gamma$          &0.47$\pm$0.02&0.47$\pm$0.05&0.47$\pm$0.05&0.47$\pm$0.02&0.47$\pm$0.02&0.48$\pm$0.04&0.49$\pm$0.04\\
4363 [O {\sc iii}]      &0.03$\pm$0.01&~~...&~~...&0.05$\pm$0.01&0.03$\pm$0.01&~~...&~~~~...\\
4861 H$\beta$           &1.00$\pm$0.03&1.00$\pm$0.04&1.00$\pm$0.04&1.00$\pm$0.03&1.00$\pm$0.03&1.00$\pm$0.04&1.00$\pm$0.04\\
4959 [O {\sc iii}]      &1.25$\pm$0.03&1.00$\pm$0.04&0.81$\pm$0.03&1.58$\pm$0.04&1.37$\pm$0.03&1.10$\pm$0.04&1.48$\pm$0.05\\
5007 [O {\sc iii}]      &3.73$\pm$0.08&2.88$\pm$0.07&2.43$\pm$0.07&4.82$\pm$0.10&4.10$\pm$0.09&3.44$\pm$0.10&4.38$\pm$0.11\\
$C$(H$\beta$)(MW)       &  0.03&  0.03&  0.04&  0.05&  0.04&  0.06&  0.02\\
$C$(H$\beta$)(int)      &  0.08&  0.08&  0.14&  0.10&  0.48&  0.00&  0.00\\
EW$_{\rm abs}$          &  2.38&  3.92&  1.96&  0.00&  0.00&  2.30&  0.00\\
EW(H$\beta$)            & 59.44& 45.14& 32.58&126.70& 95.33& 32.56& 46.95\\
$F$(H$\beta$)$^a$       & 30.57&  5.86&  8.53& 20.68& 23.44&  6.38&  8.60\\ \hline
 &J1616$+$2057 &J1716$+$2744 &J1716$+$2744 &J2145$+$0040 &J2208$-$0106 &J2211$+$0114 &J2334$-$0046 \\ \hline
2797 Mg {\sc ii}        &~~...&0.39$\pm$0.05&0.44$\pm$0.05&0.57$\pm$0.08&~~...&0.24$\pm$0.08&0.40$\pm$0.07\\
2804 Mg {\sc ii}        &~~...&0.34$\pm$0.05&0.28$\pm$0.05&0.33$\pm$0.08&~~...&0.24$\pm$0.08&0.31$\pm$0.06\\
3727 [O {\sc ii}]       &1.57$\pm$0.06&3.12$\pm$0.08&3.42$\pm$0.09&3.15$\pm$0.13&2.28$\pm$0.09&3.13$\pm$0.17&2.48$\pm$0.12\\
3868 [Ne {\sc iii}]     &0.23$\pm$0.02&0.32$\pm$0.02&0.36$\pm$0.02&0.28$\pm$0.04&0.30$\pm$0.03&0.30$\pm$0.06&0.39$\pm$0.05\\
4101 H$\delta$          &0.26$\pm$0.03&0.27$\pm$0.02&0.28$\pm$0.02&0.28$\pm$0.04&0.24$\pm$0.03&0.24$\pm$0.07&0.24$\pm$0.05\\
4340 H$\gamma$          &0.47$\pm$0.02&0.48$\pm$0.02&0.48$\pm$0.02&0.48$\pm$0.04&0.48$\pm$0.03&0.48$\pm$0.06&0.47$\pm$0.04\\
4363 [O {\sc iii}]      &0.03$\pm$0.01&~~...&~~...&~~...&~~...&~~...&~~~~...\\
4861 H$\beta$           &1.00$\pm$0.03&1.00$\pm$0.02&1.00$\pm$0.02&1.00$\pm$0.04&1.00$\pm$0.04&1.00$\pm$0.06&1.00$\pm$0.04\\
4959 [O {\sc iii}]      &1.08$\pm$0.03&1.05$\pm$0.02&1.15$\pm$0.03&0.96$\pm$0.03&1.32$\pm$0.04&0.82$\pm$0.05&1.34$\pm$0.05\\
5007 [O {\sc iii}]      &3.36$\pm$0.08&3.14$\pm$0.07&3.33$\pm$0.07&2.82$\pm$0.07&3.90$\pm$0.10&2.36$\pm$0.08&3.86$\pm$0.10\\
$C$(H$\beta$)(MW)       &  0.11&  0.07&  0.07&  0.11&  0.13&  0.05&  0.05\\
$C$(H$\beta$)(int)      &  0.13&  0.50&  0.50&  0.50&  0.00&  0.00&  0.00\\
EW$_{\rm abs}$          &  4.99&  0.92&  0.00&  1.64&  1.95&  1.40&  2.03\\
EW(H$\beta$)            & 91.08& 44.37& 44.03& 66.01& 71.18& 28.02& 54.73\\
$F$(H$\beta$)$^a$       & 12.94& 45.43& 44.73&  9.18&  7.37&  5.12&  5.30\\ \hline
 &J2348$-$0041 &J2352$+$0025 \\ \hline
2797 Mg {\sc ii}        &~~...&0.27$\pm$0.08\\
2804 Mg {\sc ii}        &~~...&0.25$\pm$0.09\\
3727 [O {\sc ii}]       &2.27$\pm$0.09&2.94$\pm$0.12\\
3868 [Ne {\sc iii}]     &0.36$\pm$0.03&0.41$\pm$0.04\\
4101 H$\delta$          &0.29$\pm$0.03&0.29$\pm$0.04\\
4340 H$\gamma$          &0.47$\pm$0.03&0.47$\pm$0.03\\
4363 [O {\sc iii}]      &0.05$\pm$0.01&0.04$\pm$0.02\\
4861 H$\beta$           &1.00$\pm$0.03&1.00$\pm$0.04\\
4959 [O {\sc iii}]      &1.32$\pm$0.03&1.34$\pm$0.04\\
5007 [O {\sc iii}]      &4.06$\pm$0.09&3.97$\pm$0.10\\
$C$(H$\beta$)(MW)       &  0.05&  0.05\\
$C$(H$\beta$)(int)      &  0.47&  0.50\\
EW$_{\rm abs}$          &  0.00&  2.09\\
EW(H$\beta$)            &100.90& 64.51\\
$F$(H$\beta$)$^a$       & 16.88&  8.01\\ \hline
 \multicolumn{8}{l}{$^a$Uncorrected H$\beta$ flux in 10$^{-16}$ erg s$^{-1}$cm$^{-2}$}
 \end{longtable}


\begin{thebibliography}{}

\bibitem[Aller(1984)]{A84} Aller, L. H. 1984, Physics of Thermal Gaseous
Nebulae (Dordrecht: Reidel)

\bibitem[Asplund et al.(2005)]{Asplund2005} Asplund, M., Grevesse, N., \&
Sauval, A.~J. 2005, ASP Conference Series, 336, 25

\bibitem[Asplund et al.(2009)]{A09} Asplund, M., Grevesse, 
N., Sauval, A.~J., \& Scott, P. 2009, ARAA, 47, 481

\bibitem[Barlow et al.(2003)]{Barlow2003} Barlow, M. J., Liu, X.-W., 
P\'equignot, D., Storey, P. J., Tsamis, Y. G., \& Morisset, C. 
2003, in Planetary nebulae: their evolution and role in the Universe, IAU
Symposium, 209, 373

\bibitem[Bordoloi et al.(2011)]{Bordoloi2011} Bordoloi, R., et al. 2011, 
\apj, 743, 10

\bibitem[Bouch\'e et al.(2006)]{Bouche2006} Bouch\'e, N., 
Murphy, M. T., P\'eroux, C., Csabai, I. \& Wild, V. 2006,
\mnras, 371, 495

\bibitem[Bruzual \& Charlot(2003)]{BC03} Bruzual, G., \& Charlot, S.
2003, \mnras, 344, 1000

\bibitem[Cardelli et al.(1989)]{C89} Cardelli, J. A., Clayton, G. C., \&
Mathis, J. S. 1989, \apj, 345, 245

\bibitem[Davey et al.(2000)]{Davey2000} Davey, A. R., Storey, P. J., 
\& Kisielius, R. 2000, \aaps, 142, 85

\bibitem[Dessauges-Zavadsky et al.(2003)]{DZavadsky2003} Dessauges-Zavadsky,
M., P\'eroux, C., Kim, T.-S., D'Odorico, S., \& McMahon, R. G. 2003, \mnras,
345, 447

\bibitem[Dessauges-Zavadsky et al.(2004)]{DZavadsky2004} Dessauges-Zavadsky,
M., Calura, F., Prochaska, J. X. D'Odorico, S., \& Matteucci, F.  2004, 
\aap, 416, 79

\bibitem[Dessauges-Zavadsky et al.(2006)]{DZavadsky2006} Dessauges-Zavadsky,
M., Prochaska, J. X. D'Odorico, S., Calura, F., \& Matteucci, F.  2006, 
\aap, 445, 93

\bibitem[Dessauges-Zavadsky et al.(2007)]{DZavadsky2007} Dessauges-Zavadsky,
M., Calura, F., Prochaska, J. X. D'Odorico, S., \& Matteucci, F. 2007,
\aap, 470, 431

\bibitem[Dinerstein et al.(2012)]{Dinerstein2012} Dinerstein, H. L., 
Prasla, F., \& Speck, A. K. 2012, American Astronomical 
Society, AAS Meeting No 219, No 343.01

\bibitem[Esteban et al.(2009)]{Esteban2009} Esteban, C., Bresolin, F., 
Peimbert, M., et al. 2009, \apj, 700, 654

\bibitem[Esteban et al.(2004)]{Esteban2004} Esteban, C., Peimbert, M., 
Garc\'ia,-Rojas, J., Ruiz, M. T., Peimbert, A., \& Rodr\'iguez, M. 2004, 
\mnras, 355, 229

\bibitem[Ferland et al.(1998)]{F98} Ferland, G. J., Korista, K. T.,
Verner, D. A., Ferguson, J. W., Kingdon, J. B., \& Verner, E. M. 1998,
\pasp, 110, 761

\bibitem[Grevesse \& Sauval(1998)]{Grevesse1998} Grevesse, N., \&
Sauval, A. J. 1998, Space Science Reviews, 85, 161

\bibitem[Guseva et al.(2011)]{Guseva2011} Guseva, N. G., Izotov, Y. I.,
 Stasi\'nska, G., Fricke, K. J., Henkel, C., \& Papaderos, P. 2011, 
\aap, 529, 149

\bibitem[Hillier \& Miller(1998)]{HM98} Hillier, D. J., \& Miller, D. L. 
1998, \apj, 496, 407

\bibitem[Hillier et al.(2001)]{H01} Hillier, D. J., Davidson, K., 
Ishibashi, K., \& Gull, T. 2001, \apj, 553, 837

\bibitem[Izotov \& Thuan(2007)]{IT07} Izotov, Y. I., \& Thuan, T. X. 2007, 
\apj, 665, 1115

\bibitem[Izotov et al.(1994)]{ITL94} Izotov, Y. I., Thuan, T. X., \& 
Lipovetsky, V. A. 1994, \apj, 435, 647 

\bibitem[Izotov et al.(1997)]{ITL97} Izotov, Y. I., Thuan, T. X., \& 
Lipovetsky, V. A. 1997, \apjs, 108, 1

\bibitem[Izotov et al.(2006)]{I06} Izotov, Y. I., Stasi\'nska, G., 
Meynet, G., Guseva, N. G., \& Thuan, T. X. 2006, \aap, 448, 955

\bibitem[Jenkins(2009)]{Jenkins2009} Jenkins, E. B. 2009, 
\apj, 700, 1299

\bibitem[Lebouteiller et al.(2008)]{Lebouteiller2008} Lebouteiller, V., 
Bernard-Salas, J., Brandl, B., et al. 2008, \apj, 680, 398


\bibitem[Leitherer et al.(1999)]{L99} Leitherer, C., Schaerer, D.,
Goldader, J. D., Gonzalez Delgado, R. M., Robert, C., Kune D. F.,
de Mello, D. F., Devost, D., \& Heckman, T. M. 1999, \apjs, 123, 3

\bibitem[Lilly et al.(2007)]{Lilly2007} Lilly, S. J., et al. 2007, 
\apjs, 172, 70

\bibitem[Lundgren et al.(2009)]{Lundgren2009} Lundgren, B. F.,
Brunner, R. J., Donald G. York, D. G., et al. 2009, \apj, 698, 819

\bibitem[Mendoza(1983)]{M83} Mendoza, C. 1983, in Planetary Nebulae,
ed. D. R. Flower, Dordrecht, D. Reidel Publishing Co., p.143

\bibitem[Meynet et al.(1994)]{M94} Meynet, G., Maeder, A., Schaller, G.,
Schaerer, \& D. Charbonnel, C. 1994, \aaps, 103, 97

\bibitem[Middlemass(1988)]{M88} Middlemass, D. 1988, \mnras, 231, 1025

\bibitem[Molaro et al.(1986)]{Molaro1986} Molaro, P., Vladilo, G., \&
Beckman, J. E. 1986, \aj, 161, 339

\bibitem[Murray(1983)]{Murray1983} Murray, M. J. 1983, Irish Astronomical 
Journal, 16, 41

\bibitem[Nestor et al.(2005)]{Nestor2005} Nestor, D. B., Turnshek, D. A., 
\& Rao, S. M. 2005, \apj, 628, 637

\bibitem[Oegerle et al.(1982)]{Oegerle} Oegerle, W. R., Kondo, Y., Stencel, 
R. E., \& Weiler, E. J. 1982, \apj, 252, 302

\bibitem[Pauldrach et al.(2001)]{P01} Pauldrach, A. W. A., Hoffman, T. L., \& 
Lennon, M. 2001, \aap, 375, 161

\bibitem[Peimbert et al.(2005)]{Peimbert2005} Peimbert, A., Peimbert, M., 
\& Ruiz, M. T. 2005, \apj, 634, 1056

\bibitem[Peimbert \& Peimbert(2010)]{PP10} Peimbert, A., \& Peimbert, M.
2010, \apj, 724, 791

\bibitem[P\'equignot \& Stasi\'nska(1980)]{PS80} P\'equignot, D., \& 
Stasi\'nska, G. 1980, \aap, 81, 121

\bibitem[Perinotto \& Patriarchi(1980)]{PP80} Perinotto, M., \& Patriarchi, P.
1980, \apj, 235, L13

\bibitem[Prochaska et al. (2007a)]{Prochaska2007a} Prochaska, J. X.,
Chen, H.-W., Bloom, J. S., et al. 2007, \apjs, 168, 231

\bibitem[Prochaska et al. (2007)]{Prochaska2007} Prochaska, J. X.,
Wolfe, A. M., Howk, J. C., Gawiser, E., Burles, S. M., \& Cooke, J. 2007,
\apjs, 171, 29

\bibitem[Prochter et al. (2006)]{Prochter2006} Prochter, G. E., 
Prochaska, J. X., \& Burles, S. M. 2006, \apj, 639, 766

\bibitem[Quider et al. (2011)]{Quider2011} Quider, A. M., Nestor, D. B., 
Turnshek, D. A., et al. 2011, \aj, 141, 137

\bibitem[Redfield \& Linsky (2004)]{Redfield2004} Redfield, S., \& Linsky, J. L.
2004, \apj, 602, 776

\bibitem[Rodr\'iguez \& Rubin (2005)]{Rodriguez2005} Rodr\'iguez, M., 
\& Rubin, R. H. 2005, \apj, 626, 900

\bibitem[Sargent et al. (1988)]{Sargent1988} Sargent, W. L. W., 
Steidel, C. C., \& Boksenberg, A. 1988, \apj, 334, 22

\bibitem[Savage et al. (1992)]{Savage1992} Savage, B. G., Cardelli, J. A.,
\& Sofia, U. J. 1992, \apj, 401, 706

\bibitem[Sigut \& Pradhan(1995)]{SP95} Sigut, T. A. A., \& Pradhan, A. K.
1995, J.Phys.B: At.Mol.Opt.Phys. 28, 4879

\bibitem[Snow \& Witt(1996)]{Snow1996} Snow, T. P., \& Witt, A. N. 1996,
\apj, 468, L65

\bibitem[Stasi\'nska \& Izotov(2003)]{SI03} Stasi\'nska, G., \&
Izotov, Y. I. 2003, \aap, 397, 71

\bibitem[Steidel \& Sargent(1992)]{Steidel1992} Steidel, C. C., \& 
Sargent, W. L. W. 1992, \apjs, 80, 1

\bibitem[Thuan et al.(1995)]{TIL95} Thuan, T. X., 
Izotov, Y. I., \& Lipovetsky, V. A. 1995, \apj, 445, 108

\bibitem[Tsamis \& P\'equignot(2005)]{Tsamis2005} Tsamis, Y. G., \& 
P\'equignot, D. 2005, \mnras, 364, 687

\bibitem[Wang \& Liu(2007)]{Wang2007} Wang, W., \& Liu, X.-W. 2007,
\mnras, 381, 669

\bibitem[Welty et al.(1999)]{Welty1999} Welty, D. E., Hobbs, L. M.,
Lauroesch, J. T., et al. 1999, \apjs, 124, 465

\bibitem[York et al.(2000)]{Y00} York, D. G., Adelman, J., 
Anderson, J. E., et al. 2000, \aj, 120, 1579

\bibitem[York et al.(2006)]{York2006} York, D. G., Khare, P., Vanden Berk, D., 
et al. 2006, MNRAS, 367, 945

\bibitem[Zhu \& M\'enard(2012)]{Zhu2012} Zhu, G., \& M\'enard, B. 2012,
\apj, in press, preprint arXiv:astro-ph1211.6215v1

\end{thebibliography}
\end{document}